\documentclass[aps,prb,twocolumn,showpacs,amsmath,amssymb,floatfix]{revtex4-1}

\usepackage{graphicx}
\usepackage{amsmath}
\usepackage{epsfig, float}
\usepackage{natbib}
\usepackage{bm}
\usepackage{latexsym}

\begin{document}

\title{The Franz-Keldysh effect revisited: Electroabsorption including interband coupling and excitonic effects}
\author{Federico Duque-Gomez}
\email{fduque@physics.utoronto.ca}
\author{J. E. Sipe}
\affiliation{Department of Physics, University of Toronto, Toronto, Ontario, Canada M5S1A7}
\date{\today}

\begin{abstract}
	We study the linear optical absorption of bulk semiconductors in the presence of a homogeneous constant (dc) electric field with an approach suitable for including excitonic effects while working with many-band models. The absorption coefficient is calculated from the time evolution of the interband polarization excited by an optical pulse. We apply the formalism to a numerical calculation for GaAs using a 14-band $\textbf{k} \cdot \textbf{p}$ model, which allows us to properly include interband coupling, and the exchange self-energy to account for the excitonic effects due to the electron-hole interaction. The Coulomb interaction enhances the features of the absorption coefficient captured by the $\textbf{k} \cdot \textbf{p}$ model. We consider the dependence of the enhancement on the strength of the dc field and the polarization of the optical field.
\end{abstract}


\maketitle

\section{INTRODUCTION}

The photoexcitation of carriers in the presence of an applied electric field was first considered by Franz\cite{Franz58} and Keldysh\cite{Keldysh58} in a calculation of the optical absorption coefficient. In the presence of a homogeneous electric field, the absorption spectrum of bulk semiconductors develops an exponential tail below the band gap and oscillations above it.\cite{Moss61, Frova66} These features became known as the Franz-Keldysh effect, and were understood very early using the effective mass approximation, where the absorption coefficient can be calculated in terms of Airy functions.\cite{Callaway63,Tharmalingam63,Penchina65,Aspnes66,Aspnes67}\\ 

For realistic band structures it is necessary to account for anisotropy, non-parabolicity and degeneracies,\cite{Keldysh70, Bottka71, Aspnes74} which require proper consideration of the interband coupling beyond the effective mass approximation.\cite{Aspnes75} For this purpose, $\mathbf{k} \cdot \mathbf{p}$ models\cite{Kane75, Pfeffer96} are particularly useful, and have been successfully applied in studies of the Franz-Keldysh effect,\cite{Enderlein75, Hader97, Wahlstrand10} providing detailed descriptions of the relation between the band structure near the $\Gamma$ point and the features of the absorption spectrum at the independent particle level. The Coulomb interaction, however, modifies significantly the optical properties due to the formation of correlated electron-hole pairs, and needs to be included. Initial efforts to add these excitonic effects, both numerically and semianalytically, were made within the effective mass approximation;\cite{Ralph68, Dow70, Blossey70, Dow71, Fauchier74, Merkulov74} this simple approximation is still used to analyze experimental data.\cite{Komkov09} Rees provided a more complete method to compare theory and experiment,\cite{Rees68, Galbraith93} including not only the electron-hole interaction but also lattice defects, band anisotropy, lifetime broadening and surface effects. The starting point in Rees's approach is the absorption spectrum without a dc field, which can then be convoluted with Airy functions that provide the field-induced effects.\\

In the present work we study the Franz-Keldysh effect considering both interband coupling and the electron-hole interaction, with an approach that is suitable for realistic band structure models. We apply the formalism to a 14-band $\mathbf{k} \cdot \mathbf{p}$ model for bulk GaAs, and compare the results with a parabolic two-band model in the effective mass approximation. The strategy is to find a gauge-independent dynamical equation for the density matrix of the system, which contains the interband polarization. For simplicity, we consider the initial condition of a crystal with valence bands fully occupied  and conduction bands completely empty, without any initial correlations between bands; afterwards, the system is excited by an optical pulse that drives the interband polarization. In linear response we can use the interband polarization to derive an expression for the absorption coefficient. Some care is necessary to account for the interband coupling due to the dc field; such coupling is treated non-perturbatively, but Zener tunneling is neglected. We follow the approach of Wahlstrand and Sipe,\cite{Wahlstrand10} reformulated appropriately for the calculation of the interband polarization from gauge-independent equations; we keep some of their notation when convenient. The Coulomb interaction is included through the exchange contribution of the Hartee-Fock self-energy, an approximation that is commonly used.\cite{HaugKoch04} \\

As expected, we observe that the Coulomb interaction enhances the features of the absorption coefficient; such enhancement is qualitatively the same in the two-band and 14-band models, but the latter allows us to address features that arise from more realistic band models, such as the dependence of the absorption coefficient on the optical polarization. We also analyze the time evolution of the interband polarization for pulsed optical excitation, which provides an useful insight into the system and the process of optical injection in the conduction bands.\footnote{A similar approach has already been used in semiconductor quantum wires. For example, see: S. Hughes and D. S. Citrin, Phys. Rev. Lett. \textbf{84}, 4228 (2000).}\\

In Sec.~\ref{S:THEORETICAL_FRAMEWORK} we present the formalism that leads to the dynamical equation for the interband polarization, and show how the absorption coefficient is calculated from the interband polarization. In Sec.~\ref{S:Calculations} we discuss the diverse quantities required as input for the calculation in the two models mentioned before; the methods, however, can be easily applied to any other type of $\mathbf{k} \cdot \mathbf{p}$ calculation. We also illustrate the time evolution of the interband polarization with some examples in Sec.~\ref{S:Calculations}, before presenting the results for the absorption coefficient in Sec.~\ref{S:Results}. The absorption coefficient is calculated with and without Coulomb interaction and for different dc fields. Finally, we present some conclusions in Sec.~\ref{S:Conclusion}.\\ 

\section{THEORETICAL FRAMEWORK  \label{S:THEORETICAL_FRAMEWORK}}

The goal of this section is to derive expressions for the interband polarization, and the absorption coefficient, for a system of charges excited by an optical pulse in the presence of a crystal potential $V(\mathbf{x})$ and an applied dc electric field. We describe the electromagnetic fields classically  in the long-wavelength limit with a uniform electric field, $\mathbf{E}(t)=\mathbf{E}^{\text{dc}}(t) + \mathbf{E}^{\text{opt}}(t)$. The optical pulse, $\mathbf{E}^{\text{opt}}(t)$, is centered at $t=0$ and its amplitude is effectively zero outside a given time range $\left[t_{\text{o}}, t_{\text{f}}\right]$. The dc field, $\mathbf{E}^{\text{dc}}(t)$, is applied at some initial time $t_{\text{init}} \ll t_o$ and left constant afterwards.\\

We begin in Sec.~\ref{S:LesserGreenFunction} with the dynamical equation of the density matrix for the system in reciprocal space. Treating the optical field perturbatively in Sec.~\ref{S:PerturbativeSolutionOpticalField}, we derive an equation for the interband polarization to first order in this field. The effects of the dc field are discussed in Sec.~\ref{S:InterbandCouplingDCfield}; the interband coupling due to the dc field is included non-perturbatively, but we neglect Zener tunneling. Finally, in Sec.~\ref{S:AbsorptionCoefficient} we use the interband polarization to find an expression for the absorption coefficient.


\subsection{Lesser Green function \label{S:LesserGreenFunction}} 

The populations and correlations necessary for studying the Franz-Keldysh effect are contained in the lesser Green function
\begin{equation}
	\mathcal{G}^{<}(\mathbf{x}_1, t_1; \mathbf{x}_2, t_2) \equiv -\frac{1}{i\hbar} \left< \psi^{\dagger}(\mathbf{x}_2,t_2) \psi(\mathbf{x}_1,t_1) \right>, \label{E:DefLesserGreen}
\end{equation}
where $\psi(\mathbf{x},t)$ is the electron field operator. The simplest approximation to include excitonic effects due to the electron-hole interaction is through the exchange contribution of the Hartree-Fock self-energy, 
\begin{multline}
	\Sigma^{\text{ex}}(\mathbf{x}_1,t_1;\mathbf{x}_2,t_2) \equiv \\
	i\hbar \, U(\mathbf{x}_1-\mathbf{x}_2) \delta(t_1-t_2) \mathcal{G}^{<}(\mathbf{x}_1,t_1;\mathbf{x}_2,t_2), \label{E:ExchangeSelfEnergy}
\end{multline}
where $U(\mathbf{x})$ is the Coulomb potential.\cite{HaugKoch04} For the calculations in this paper, we consider the screening effects at the level of the polarizability of the valence electrons and of the lattice, captured in a \emph{background dielectric constant} (see Eq.~\eqref{E:BareCoulomb} in Sec.~\ref{S:Time_ev_InterPol});\cite{HaugKoch04} a \emph{static screening} is also discussed briefly in Sec.~\ref{S:Results}. Within this simple Hartree-Fock treatment, it is enough to consider the density matrix of the system for calculating the interband polarization and the density of excited carriers. The density matrix is essentially Eq.~\eqref{E:DefLesserGreen} in the equal-time limit, $\mathcal{G}^{<}(\mathbf{x}_1, t; \mathbf{x}_2, t)$, where the average time $(t_1+t_2)/2 \rightarrow t$ and the time difference $t_1-t_2 \rightarrow 0$.\\

We are interested in a situation where the only applied field is a uniform electric field $\mathbf{E}(t)$, which can be introduced through a uniform vector potential $\mathbf{A}(t)$. In this case it is convenient to expand $\psi(\mathbf{x}, t)$ in terms of field operators $a_n^{\prime}(\mathbf{k}, t)$ for the usual Bloch states $\phi_{n \mathbf{k}}(\mathbf{x})$ modified by an additional phase,
\begin{equation}
	\phi_{n \mathbf{k}}^{\prime}(\mathbf{x},t) \equiv \phi_{n \mathbf{k}}(\mathbf{x}) e^{i \frac{e}{\hbar c} \mathbf{A}(t) \cdot \mathbf{x}}. \notag
\end{equation}
The Bloch states $\phi_{n \mathbf{k}}(\mathbf{x})$, labeled by band index $n$ and crystal wavevector $\mathbf{k}$ are solutions of the eigenvalue problem
\begin{equation}
	\mathcal{H}_{o} \phi_{n \mathbf{k}}(\mathbf{x}) = \hbar \omega_{n}(\mathbf{k}) \phi_{n \mathbf{k}}(\mathbf{x}), \notag
\end{equation}
for the unperturbed Hamiltonian, of which the simplest form is 
\begin{equation}
	\mathcal{H}_{o} \equiv \frac{1}{2m} \left( \frac{\hbar}{i} \nabla \right)^2 + V(\mathbf{x}). \label{E:DefUnperturbedHamiltonian}
\end{equation}
They have the form
\begin{equation}
	\phi_{n \mathbf{k}}(\mathbf{x}) = \frac{1}{(2 \pi)^{3/2}} u_{n \mathbf{k}}(\mathbf{x}) e^{i \mathbf{k} \cdot \mathbf{x}}, \notag
\end{equation}
where $ u_{n \mathbf{k}}(\mathbf{x})$ is a function with the periodicity of the lattice potential $V(\mathbf{x})$. In practice we use band structures that include spin-orbit coupling and in principle other relativistic corrections. For situations where only points and lines of degeneracy are present, the Bloch functions can be constructed to be periodic in reciprocal space and therefore well-defined for all $\mathbf{k}$, except for degeneracy points.\cite{Lax74,  Gharamani93} Note that the modified Bloch functions form a complete set of orthonormal states; in this representation, $ \mathcal{G}^{<}(\mathbf{x}_1,t;\mathbf{x}_2,t)$ becomes
\begin{equation}
	G_{n_1 n_2}^{<}(\mathbf{k}_1, t; \mathbf{k}_2, t) \equiv -\frac{1}{i\hbar} \left< a^{\prime \dagger}_{n_2}(\mathbf{k}_2,t) a_{n_1}^{\prime}(\mathbf{k}_1,t) \right>. \notag
\end{equation}\\ 
Within our approximations, only the relative position between the electron and hole is relevant (see Appendix); therefore, we look for solutions of the form
\begin{equation}
	G_{n_1 n_2}^{<}(\mathbf{k}_1, t; \mathbf{k}_2, t) = \delta(\mathbf{k}_1 - \mathbf{k}_2)G_{n_1 n_2}^{<}(\mathbf{k}_1, t). \label{E:GreenFuncBloch}
\end{equation}
The Green function $G_{n_1 n_2}^{<}(\mathbf{k}, t)$ can be considered as the $(n_1,n_2)$ element of a matrix $G^{<}(\mathbf{k}, t)$ labeled by the crystal wavevector $\mathbf{k}$, which is associated with the relative position of the electron-hole pair. The time evolution of such matrix is governed by the dynamical equation
\begin{multline}
	i \hbar \frac{\partial}{\partial t} G^{<}(\mathbf{k},t) + e \mathbf{E}(t) \cdot i \nabla_{\mathbf{k}}  G^{<}(\mathbf{k},t) \\ 
	- \left[ \beta(\mathbf{k},t) + \Sigma(\mathbf{k},t), G^{<} (\mathbf{k},t) \right] = 0, \label{E:DysonEqBloch}
\end{multline}
where $e=-|e|$ is the electron charge. Note that this equation is formulated in terms of the electric field $\mathbf{E}(t)$ instead of the vector potential $\mathbf{A}(t)$. In the Appendix we derive Eq.~\eqref{E:DysonEqBloch} from a gauge-independent formalism using a basis of Wannier functions modified by the Peierls phase,\cite{KitaYamashita08} but it can also be obtained from the usual application of the length gauge.\cite{HaugKoch04, Virk07} The two energy terms in the commutator of Eq.~\eqref{E:DysonEqBloch} correspond to the single-particle energy, $\beta(\mathbf{k},t)$, and the self-energy, $ \Sigma(\mathbf{k},t)$; at each $\mathbf{k}$ they are matrices labeled by the band indices. The former is given by 
\begin{equation}
	\beta_{n_1 n_2}(\mathbf{k},t) \equiv \delta_{n_1 n_2} \hbar \omega_{n_1}(\mathbf{k}) - e \mathbf{E}(t) \cdot \boldsymbol \xi_{n_1 n_2}(\mathbf{k}) \notag
\end{equation}
with the matrix elements \cite{Lax74}
\begin{equation}
	\boldsymbol \xi_{n_1 n_2}(\mathbf{k}) \equiv  \frac{1}{\Omega_{\text{cell}}} \int_{\text{cell}} u_{n_1 \mathbf{k}}^{\ast}(\mathbf{x}) i \nabla_{\mathbf{k}} u_{n_2 \mathbf{k}}(\mathbf{x}) d \mathbf{x}, \label{E:DefLaxConnection}
\end{equation}
where the integration is over the unit cell of volume $\Omega_{\text{cell}}$; the diagonal element $\boldsymbol \xi_{n n}(\mathbf{k})$ is known as the \emph{Berry connection}.\cite{Berry84, Price12} In this Bloch state representation, the matrix representing the exchange self-energy, Eq.~\eqref{E:ExchangeSelfEnergy}, is
\begin{multline}
	\Sigma(\mathbf{k},t) = i\hbar \int \frac{d\mathbf{q}}{(2 \pi)^3} U(\mathbf{q}) \\
	\times \Delta(\mathbf{k}, \mathbf{k}-\mathbf{q}) G^{<}(\mathbf{k}-\mathbf{q},t)  \Delta^{\dagger}(\mathbf{k}, \mathbf{k}-\mathbf{q}), \label{E:ExchangeSelfEnergyBloch}
\end{multline}
where $U(\mathbf{q})$ is the Fourier transform of the Coulomb potential,
\begin{equation}
	U(\mathbf{q}) = \int U(\mathbf{x}) e^{-i\mathbf{q} \cdot \mathbf{x}} d\mathbf{x}, \notag
\end{equation}
and the overlap matrix $\Delta(\mathbf{k}_1, \mathbf{k}_2)$ has elements \cite{LewYanVoon09}
\begin{equation}
	\Delta_{n_1 n_2}(\mathbf{k}_1, \mathbf{k}_2) \equiv \frac{1}{\Omega_{\text{cell}}} \int_{\text{cell}} u_{n_1 \mathbf{k}_1}^{\ast}(\mathbf{x}) u_{n_2 \mathbf{k}_2}(\mathbf{x}) d\mathbf{x}. \label{E:OverlapBlochCoulomb}
\end{equation}\\


\subsection{Perturbative solution in the optical field \label{S:PerturbativeSolutionOpticalField}}

In order to solve Eq.~\eqref{E:DysonEqBloch} for $\mathbf{E}(t)=\mathbf{E}^{\text{dc}}(t) + \mathbf{E}^{\text{opt}}(t)$, we use an expansion of the lesser Green function in powers of the optical field,
\begin{equation}
	G^{<}(\mathbf{k},t) = G^{<(0)}(\mathbf{k},t) + G^{<(1)}(\mathbf{k},t) + ... \label{E:PerturbationExpansion}
\end{equation}
Accordingly, the single-particle energy can be decomposed as $\beta(\mathbf{k},t) = \beta^{(0)}(\mathbf{k},t) + \beta^{(1)}(\mathbf{k},t)$, with
\begin{equation}
	\beta_{n_1 n_2}^{(0)}(\mathbf{k},t) \equiv \delta_{n_1 n_2} \hbar \omega_{n_1}(\mathbf{k}) - e \mathbf{E}^{\text{dc}}(t) \cdot \boldsymbol \xi_{n_1 n_2}(\mathbf{k}) \label{E:SingleParticleEnergyZero}
\end{equation}
and
\begin{equation}
	\beta_{n_1 n_2}^{(1)}(\mathbf{k},t) \equiv - e \mathbf{E}^{\text{opt}}(t) \cdot \boldsymbol \xi_{n_1 n_2}(\mathbf{k}), \notag
\end{equation}
where we associate $\mathbf{E}^{\text{opt}}(t)$ with the Maxwell field in the medium. The exchange self-energy is also expanded in powers of the optical field according to which term of Eq.~\eqref{E:PerturbationExpansion} is used in Eq.~\eqref{E:ExchangeSelfEnergyBloch}. The zeroth order self-energy, $\Sigma^{(0)}(\mathbf{k},t)$, essentially renormalizes the band energies and matrix elements.\cite{HaugKoch04}  This effect can be absorbed in the band structure model, such as the $\mathbf{k} \cdot \mathbf{p}$ model that will be used in Sec.~\ref{S:Calculations}; therefore, we will not include $\Sigma^{(0)}(\mathbf{k},t)$ explicitly.\\ 

We keep the dc field to all orders but we neglect the coupling between conduction and valence bands due to the dc field; this is equivalent to neglecting Zener tunneling.\cite{Wahlstrand10} Hence, we do not include the contribution of $ - e \mathbf{E}_{\text{dc}}(t) \cdot \boldsymbol \xi_{n_1 n_2}(\mathbf{k})$ to Eq.~\eqref{E:SingleParticleEnergyZero}  when $(n_1,n_2)$ are not both conduction or both valence bands. Within this approximation, the matrix $\beta^{(0)}(\mathbf{k},t)$ takes a block diagonal form with zero matrix elements between conduction and valence bands. Matrices of this form are denoted by a tilde on top; therefore, instead of the full $\beta^{(0)}(\mathbf{k},t)$ we use $\tilde{\beta}^{(0)}(\mathbf{k},t)$. A consequence of this approximation is that $G^{<(0)}(\mathbf{k},t)$ does not change in time for an initial state with valence bands fully occupied and empty conduction bands. Thus, we have
\begin{eqnarray}
	G^{<(0)}_{n_1 n_2}(\mathbf{k},t) &=& -\frac{\delta_{n_1 n_2}}{i \hbar} \, \text{ if $n_1$ and $n_2$ are valence bands,} \notag \\ &=& 0 \, \text{ otherwise,} \notag
\end{eqnarray}
for all $t$, even after the dc field is applied. Additionally, the first order contribution, $G^{<(1)}_{n_1 n_2}(\mathbf{k},t)$, is different from zero only after the pulse starts at $t_{o}$ and if $n_1$ and $n_2$ are not both conduction or both valence bands. Matrices of this off-diagonal block form are denoted by a bar on top; thus, we can replace $G^{<(1)}(\mathbf{k},t)$ by $\bar{G}^{<(1)}(\mathbf{k},t)$. Note that $\bar{G}^{<(1)}(\mathbf{k},t)$ is essentially the interband polarization between conduction and valence bands, so we define the matrix
\begin{equation}
	\bar{\Pi}(\mathbf{k}, t) = -i\hbar \bar{G}^{< (1)}(\mathbf{k},t). \notag
\end{equation}
Using Eq.~\eqref{E:DysonEqBloch} we find that the time evolution of $\bar{\Pi}_{cv}(\mathbf{k}, t)$, the matrix element of $\bar{\Pi}(\mathbf{k}, t)$ associated with a conduction band $c$ and a valence band $v$, is governed by
\begin{multline}
	i \hbar \frac{\partial}{\partial t} \bar{\Pi}_{cv}(\mathbf{k},t) +  i e E^{\text{dc}}(t) \frac{\partial}{\partial k_{\parallel}} \bar{\Pi}_{cv}(\mathbf{k},t) \\
	- \left[ \tilde{\beta}^{(0)}(\mathbf{k},t), \bar{\Pi}(\mathbf{k},t) \right]_{cv} - \Sigma_{cv}^{(1)}(\mathbf{k},t) = \bar{\beta}^{(1)}_{cv}(\mathbf{k},t), \label{E:DysonEqBlochFirst}
\end{multline}
which has the same form as the usual semiconductor Bloch equations.\cite{HaugKoch04} Since we only need the matrix elements of $\beta^{(1)}(\mathbf{k},t)$ that couple conduction and valence bands, we use $\bar{\beta}^{(1)}(\mathbf{k},t)$. In Eq.~\eqref{E:DysonEqBlochFirst} we denote the component of $\mathbf{k}$ parallel to the dc field by $k_{\parallel}$; later, we will use $\mathbf{k}_{\perp}$ for the perpendicular component.\\


\subsection{Interband coupling due to the dc field \label{S:InterbandCouplingDCfield}}

The terms introduced in Eq.~\eqref{E:DefLaxConnection} for $n_1 \ne n_2$ are responsible for the interband coupling induced by the electric field $\mathbf{E}(t)$. We reformulate Eq.~\eqref{E:DysonEqBlochFirst} so that the interband coupling due to the dc field is calculated separately from the dynamical equation for the interband polarization. In our approach we apply a unitary matrix transformation accompanied by appropriate phase factors; both are discussed in this subsection.\\

For the matrix transformation, recall that the off-diagonal terms of $\boldsymbol \xi(\mathbf{k})$ for non-degenerate bands are given by \cite{Lax74}
\begin{equation}
	\boldsymbol \xi_{n_1 n_2}(\mathbf{k}) \equiv \frac{\mathbf{v}_{n_1 n_2}(\mathbf{k})}{i \omega_{n_1 n_2}(\mathbf{k})}, \notag
\end{equation}  
where $\omega_{n_1 n_2}(\mathbf{k}) = \omega_{n_1}(\mathbf{k})-\omega_{n_2}(\mathbf{k})$. The $\mathbf{v}_{n_1 n_2}(\mathbf{k})$ are the usual velocity matrix elements
\begin{equation}
	\mathbf{v}_{n_1 n_2}(\mathbf{k}) \equiv \frac{(2 \pi)^3}{\Omega_{\text{cell}}} \int_{\text{cell}} \phi_{n_1 \mathbf{k}}^{\ast}(\mathbf{x}) \frac{\hbar}{i m} \nabla \phi_{n_2 \mathbf{k}} (\mathbf{x}) d\mathbf{x} \notag
\end{equation}
in the Bloch representation. It is convenient to transform the $\mathbf{v}_{n_1 n_2}(\mathbf{k})$ according to
\begin{equation}
	\mathbf{V}(\mathbf{k}) \equiv e^{i \Xi(\mathbf{k})} \mathbf{v}(\mathbf{k}) e^{-i \Xi(\mathbf{k})}, \label{E:VelocityMatrixElementKP}
\end{equation}
where 
\begin{equation}
	\Xi_{n_1 n_2}(\mathbf{k}) \equiv - \delta_{n_1 n_2} \int_{0}^{k_{\parallel}} \xi_{n_1 n_1}^{\text{dc}}(\mathbf{k}_{\perp} + k_{\parallel}' \hat{\mathbf{e}}^{\text{dc}})d k_{\parallel}' \label{E:Berry_typePhase}
\end{equation}
is the phase associated with the Berry connection along the dc field direction, $ \xi_{n n}^{\text{dc}}(\mathbf{k})$. Here $\hat{\mathbf{e}}^{\text{dc}}$ is a unit vector in the direction of the dc field. The advantage of this phase transformation will be evident when we apply it to $\bar{\Pi}(\mathbf{k}, t)$ in order to remove the explicit dependence of the interband polarization on the phase of the Bloch functions fixed by $\xi_{n n}^{\text{dc}}(\mathbf{k})$. Now we can introduce the dipole moment matrix with elements
\begin{eqnarray}
	\boldsymbol \mu_{n_1 n_2}(\mathbf{k}) &\equiv& \frac{e \mathbf{V}_{n_1 n_2}(\mathbf{k})}{i \omega_{n_1 n_2}(\mathbf{k})} \, \text{ if $\omega_{n_1 n_2}(\mathbf{k}) \ne 0$,} \nonumber \\ &\equiv& 0 \, \text{ otherwise.} \label{E:DefDipoleMatrixElement}
\end{eqnarray}
As usual, the block diagonal version of this matrix, without coupling between conduction and valence bands, is denoted by $\tilde{\boldsymbol \mu} (\mathbf{k})$. The remaining matrix elements, which connect conduction and valence bands, are kept in the off-diagonal block matrix $\bar{\boldsymbol \mu} (\mathbf{k})$. In order to neglect Zener tunneling, we only keep the coupling due to $\tilde{\boldsymbol \mu} (\mathbf{k})$ and we define a block diagonal matrix $\tilde{M}(\mathbf{k})$ that satisfies the differential equation
\begin{multline}
	i e E^{\text{dc}} \frac{\partial}{\partial k_{\parallel}} \tilde{M}_{n_1 n_2}(\mathbf{k}) - \hbar \omega_{n_1 n_2}(\mathbf{k})  \tilde{M}_{n_1 n_2}(\mathbf{k}) \\
	+ \sum_{n_1'} E^{\text{dc}} \tilde{\mu}_{n_1 n_1'}^{\text{dc}}(\mathbf{k}) \tilde{M}_{n_1' n_2}(\mathbf{k}) = 0, \label{E:dcInterCouplingDiffEq}
\end{multline}
subject to the condition $\tilde{M}_{n_1 n_2}(\mathbf{k}_{\perp})=\delta_{n_1 n_2}$ on the plane with $k_{\parallel}=0$. Here $E^{\text{dc}}$ is the magnitude of the dc field once it is left constant, and $\tilde{\mu}_{n_1 n_2}^{\text{dc}}(\mathbf{k})$ is the component of $\tilde{\mu}_{n_1 n_2} (\mathbf{k})$ parallel to $\hat{\mathbf{e}}^{\text{dc}}$.\\

We define our transformed interband polarization according to 
\begin{equation}
	\bar{P}(\mathbf{k}, t) \equiv \tilde{M}^{\dagger}(\mathbf{k}) e^{i \Xi(\mathbf{k})} \bar{\Pi}(\mathbf{k}, t) e^{-i \Xi(\mathbf{k})} \tilde{M}(\mathbf{k}), \label{E:DefNewInterPol}
\end{equation} 
which combines the phase factors used previously in Eq.~\eqref{E:VelocityMatrixElementKP} and the matrix $\tilde{M}(\mathbf{k})$.  Note that the time evolution of $\bar{P}_{cv}(\mathbf{k}, t)$,
\begin{multline}
	i\hbar \frac{\partial}{\partial t}  \bar{P}_{cv}(\mathbf{k},t) - \hbar \omega_{cv}(\mathbf{k}) \bar{P}_{cv}(\mathbf{k},t) + i e E^{\text{dc}} \frac{\partial}{\partial k_{\parallel}}\bar{P}_{cv}(\mathbf{k},t) \\
	- \Sigma_{c v}^{\bar{P}} (\mathbf{k}, t) = - \mathbf{E}^{\text{opt}}(t) \cdot \bar{\boldsymbol \Theta}_{cv}(\mathbf{k}), \label{E:InterbandPolEq}
\end{multline}
is independent of the Berry connection, and the interband coupling due to the dc field is implicit in the transformed dipole moment matrix
\begin{equation}
	\bar{\boldsymbol \Theta}(\mathbf{k}) \equiv \tilde{M}^{\dagger}(\mathbf{k}) \boldsymbol \bar{\boldsymbol \mu}(\mathbf{k}) \tilde{M}(\mathbf{k}). \label{E:TransformedDipole}
\end{equation}
The Coulomb term in Eq.~\eqref{E:InterbandPolEq},
\begin{multline}
	\Sigma^{\bar{P}}(\mathbf{k}, t) = - \int \frac{d\mathbf{q}}{(2 \pi)^3} U(\mathbf{q})\\
	\times K(\mathbf{k}, \mathbf{k}-\mathbf{q}) \bar{P}(\mathbf{k}-\mathbf{q},t) K^{\dagger}(\mathbf{k}, \mathbf{k}-\mathbf{q}), \label{E:TransformedCoulombTerm}
\end{multline}
has the same structure as Eq.~\eqref{E:ExchangeSelfEnergyBloch}, but involves the overlap matrix $\Delta(\mathbf{k}_1, \mathbf{k}_2)$ transformed to
\begin{equation}
	K(\mathbf{k}_1, \mathbf{k}_2) \equiv \tilde{M}^{\dagger}(\mathbf{k}_1) e^{i \Xi(\mathbf{k}_1)} \Delta(\mathbf{k}_1, \mathbf{k}_2) e^{-i \Xi(\mathbf{k}_2)} \tilde{M}(\mathbf{k}_2), \label{E:OverlapKP}
\end{equation}
in a generalization of the transformation indicated in Eq.~\eqref{E:DefNewInterPol}.\\

For  calculating the matrices $\mathbf{V}(\mathbf{k})$ and $K(\mathbf{k}_1, \mathbf{k}_2)$ it is actually not necessary to find the Berry connections. Wahlstrand and Sipe\cite{Wahlstrand10} presented a procedure to construct such matrices dealing properly with lines and points of degeneracy;\cite{Gharamani93} we will apply this approach in Sec.~\ref{S:14bandModel}.\\


\subsection{Absorption coefficient \label{S:AbsorptionCoefficient}}

In the perturbation expansion Eq.~\eqref{E:PerturbationExpansion}, the diagonal elements of the second order term contain the lowest order contributions to the density of carriers excited to the conduction bands. The dynamical equation for $G^{<(2)}(\mathbf{k}, t)$ is derived in the usual way, substituting Eq.~\eqref{E:PerturbationExpansion} in Eq.~\eqref{E:DysonEqBloch} and keeping only second order terms. Then the density of excited carriers can be written as
\begin{multline}
	n_{\text{exc}} = \frac{1}{\hbar} \sum_{c, v} \int_{\text{BZ}} 2 \, \text{Im} \bigg[ \int_{-\infty}^{\infty} \bar{P}_{cv}(\mathbf{k},\omega) \\
	\times \left( \mathbf{E}^{\text{opt}}(\omega) \cdot \boldsymbol \bar{\boldsymbol \Theta}_{cv}(\mathbf{k}) \right)^{\ast} \frac{d\omega}{2 \pi} \bigg] \frac{d\mathbf{k}}{(2 \pi)^3}, \label{E:DensityExcited}
\end{multline}
where 
\begin{equation}
	\bar{P}_{cv}(\mathbf{k},\omega) \equiv \int_{-\infty}^{\infty} \bar{P}_{cv}(\mathbf{k},t)  e^{i \omega t} \, d t \label{E:FourierTransfTime}
\end{equation}
is the Fourier transform of the interband polarization and BZ denotes integration over the (first) Brillouin zone. The Fourier transform of the optical pulse, $\mathbf{E}^{\text{opt}}(\omega)$, is defined similarly to Eq.~\eqref{E:FourierTransfTime}. Since the interband polarization is linear in the optical field we can write 
\begin{equation}
	 \bar{P}_{cv}(\mathbf{k},\omega) = \sum_{i} \bar{X}_{cv}^{(i)}(\mathbf{k}, \omega) E^{\text{opt}}_{i}(\omega), \label{E:ResponseFunction}
\end{equation}
where $\bar{X}_{cv}^{(i)}(\mathbf{k}, \omega)$ characterizes the response of $\bar{P}_{cv}(\mathbf{k},\omega)$ to an optical field in the $i$th Cartesian direction. After using Eq.~\eqref{E:ResponseFunction} and taking the limit to a \emph{long} pulse of frequency $\omega_o$, $\mathbf{E}^{\text{opt}}(t) \rightarrow \mathbf{E}^{\mathbf{o}} e^{-i \omega_o t} + \text{c.c.}$, we rewrite Eq.~\eqref{E:DensityExcited} in the form of Fermi's Golden Rule 
\begin{equation}
	\frac{d n_{\text{exc}}}{dt} = \sum_{ij} \eta^{ij}(\omega_o) E^{o}_{i} (E^{o}_{j})^{\ast}, \notag
\end{equation}
where the tensor
\begin{multline}
	\eta^{ij}(\omega) = \frac{1}{\hbar} \sum_{c, v} \int_{\text{BZ}} \bigg[ \bar{X}_{cv}^{(i)}(\mathbf{k}, \omega) \left( \bar{\Theta}_{cv}^{j}(\mathbf{k}) \right)^{\ast} \\
	- \left( \bar{X}_{cv}^{(j)}(\mathbf{k}, \omega) \right)^{\ast} \bar{\Theta}_{cv}^{i}(\mathbf{k}) \bigg] \frac{d\mathbf{k}}{(2 \pi)^3} \label{E:AbsorptionTensor} 
\end{multline}
describes the rate of injection of carriers into the conduction bands. If both the optical and the dc field are aligned with principal axes of the dielectric tensor, the absorption coefficient becomes \cite{Yu99, Wahlstrand10} 
\begin{equation}
	\alpha(\omega) = \frac{2 \pi \hbar \omega \eta^{ii}(\omega)}{n(\omega) c}, \notag
\end{equation}
for an optical field linearly polarized along the direction $i$ and associated refractive index $n(\omega)$; in the numerical calculations discussed in Sec.~\ref{S:Results} we neglect the frequency dependence of the refractive index and use $n(\omega)=3.7$ for GaAs.\\

While the injection rate $\eta^{ij}(\omega)$ and the absorption coefficient $\alpha(\omega)$ for CW excitation at frequency $\omega$ are the quantities in which we are primarily interested, it is useful to consider the response of the crystal to pulsed excitation in the time-domain to illustrate the physics of the injection process; furthermore, a numerical evaluation of the response to a pulse is one strategy for determining  $\eta^{ij}(\omega)$. Therefore, we consider an optical pulse of \emph{finite} duration with its polarization oriented in the $i$th Cartesian direction, and calculate the interband polarization in time-domain, $\bar{P}_{cv}^{(i)}(\mathbf{k}, t)$, from Eq.~\eqref{E:InterbandPolEq}. The linear response function, $\bar{X}_{cv}^{(i)}(\mathbf{k}, \omega)$, is then found from the Fourier components according to Eq.~\eqref{E:ResponseFunction}.\\ 

We assume a pulse of the form
\begin{equation}
	E_{i}^{\text{opt}}(t) = \mathcal{E}_{i}^{\text{opt}}(t) e^{-i \omega_o t} + (\mathcal{E}_{i}^{\text{opt}}(t))^{\ast} e^{i \omega_o t}, \notag
\end{equation}
with carrier frequency $\omega_o$  and slowly-varying envelope $\mathcal{E}_{i}^{\text{opt}}(t)$. Since we are interested in frequencies $\omega_o$ near the band gap, it is convenient to write the solution of Eq.~\eqref{E:InterbandPolEq} as $\bar{P}^{(i)}_{cv}(\mathbf{k}, t) = \bar{\mathcal{P}}^{(i)}_{cv}(\mathbf{k}, t)e^{-i \omega_o t}$. In the rotating wave approximation, the function $\bar{\mathcal{P}}^{(i)}_{cv}(\mathbf{k}, t)$ satisfies 
\begin{multline}
	i\hbar \frac{\partial}{\partial t}  \bar{\mathcal{P}}_{cv}^{(i)}(\mathbf{k},t) - \hbar(\omega_{cv}(\mathbf{k})-\omega_o) \bar{\mathcal{P}}_{cv}^{(i)}(\mathbf{k},t) \\
	+ i e E^{\text{dc}} \frac{\partial}{\partial k_{\parallel}}\bar{\mathcal{P}}_{cv}^{(i)}(\mathbf{k},t) - \Sigma_{c v}^{\bar{\mathcal{P}}^{(i)}} (\mathbf{k}, t) = -\mathcal{E}^{\text{opt}}_{i}(t) \bar{\Theta}_{cv}^{i}(\mathbf{k}), \label{E:InterbandPolRWAEq}
\end{multline} 
where $\bar{\Theta}^{i}(\mathbf{k})$ denotes the $i$th Cartesian component of the transformed dipole moment matrix and the self-energy \eqref{E:TransformedCoulombTerm} is written in terms of $\bar{\mathcal{P}}^{(i)}(\mathbf{k},t)$,
\begin{multline}
	\Sigma^{\bar{\mathcal{P}}^{(i)}} (\mathbf{k}, t) = - \int \frac{d\mathbf{q}}{(2 \pi)^3} U(\mathbf{q})\\
	\times K(\mathbf{k}, \mathbf{k} - \mathbf{q}) \bar{\mathcal{P}}^{(i)}(\mathbf{k} - \mathbf{q},t) K^{\dagger}(\mathbf{k}, \mathbf{k} - \mathbf{q}). \notag
\end{multline}
We solve Eq.~\eqref{E:InterbandPolRWAEq} numerically for a convenient Gaussian envelope 
\begin{equation}
	\mathcal{E}^{\text{opt}}_{i} (t) = \frac{1}{2} \mathcal{E}^{\mathbf{o}}_{i} e^{-(t/\tau)^2}, \notag
\end{equation} 
where $\tau$ is the duration of the pulse, and extract the linear response from the Fourier transform with
\begin{equation}
	\bar{X}_{cv}^{(i)}(\mathbf{k}, \omega) = \frac{\bar{\mathcal{P}}_{cv}^{(i)}(\mathbf{k},\omega - \omega_o)}{\mathcal{E}^{\text{opt}}_{i}(\omega - \omega_o)} \notag
\end{equation}
for frequencies near $\omega_o$. The linear response function calculated this way can then be inserted in Eq.~\eqref{E:AbsorptionTensor} to find the tensor $\eta^{ij}(\omega)$.\\
 

\section{CALCULATIONS \label{S:Calculations}} 

We illustrate the type of calculations presented in Sec.~\ref{S:THEORETICAL_FRAMEWORK} for bulk GaAs in a simple parabolic two-band model\cite{HaugKoch04, Fauchier74, Komkov09} and a 14-band $\mathbf{k}\cdot \mathbf{p}$ model\cite{Pfeffer96}. We start by calculating the various matrices that capture the interband coupling in the system. The first is the velocity matrix $\mathbf{V}(\mathbf{k})$ that couples to the dc field in Eq.~\eqref{E:dcInterCouplingDiffEq} through the usual dipole moment matrix elements, and allows us to calculate the transformation matrix $\tilde{M}(\mathbf{k})$. With these matrix elements we can calculate the modified dipole moment matrix $\bar{\boldsymbol \Theta}(\mathbf{k})$, which couples to the optical field and drives the interband polarization. Additionally, the Coulomb term requires calculating the transformed overlap matrix $K(\mathbf{k}_1, \mathbf{k}_2)$. In the two band model, discussed in Sec.~\ref{S:TwoBandModel}, these matrix elements are trivial. But in the 14-band model, described in Sec.~\ref{S:14bandModel}, we need to combine the standard $\mathbf{k}\cdot \mathbf{p}$ method with the equations described in Sec.~\ref{S:InterbandCouplingDCfield}. The resulting matrices are required in the dynamical equation for the interband polarization; we discuss the approach to solve numerically such equation in Sec.~\ref{S:Time_ev_InterPol}.


\subsection{The parabolic two-band model \label{S:TwoBandModel}}

In the parabolic two-band model the energy difference between the conduction and the valence bands around the $\Gamma$ point is simply given by
\begin{equation}
	\hbar \omega_{cv}(\mathbf{k}) = \frac{\hbar^2 \mathbf{k}^2}{2 m^{\ast}_{cv}} + E_{g}, \notag
\end{equation}
where $m^{\ast}_{cv}$ is the reduced effective mass of the electron-hole pair and $E_{g}$ is the energy gap. The $u_{n \mathbf{k}}(\mathbf{x})$ and the $\mathbf{v}_{cv}(\mathbf{k})$ are fixed to the value at the $\Gamma$ point. Therefore, 
\begin{equation}
	\mathbf{V}_{cv}(\mathbf{k}) = \mathbf{v}_{cv}(\boldsymbol \Gamma). \label{E:VelMatElTwoBand}
\end{equation}
Similarly, the interband coupling due to the dc field, Eq.~\eqref{E:dcInterCouplingDiffEq}, and the transformed overlap matrices, Eq.~\eqref{E:OverlapKP}, are uniform in $\mathbf{k}$,
\begin{equation}
	\tilde{M}_{n_1 n_2}(\mathbf{k}) = K_{n_1 n_2}(\mathbf{k}_1, \mathbf{k}_2) = \delta_{n_1 n_2}, \notag
\end{equation}
where $n_1$ and $n_2$ can only be either $c$ or $v$. This model has been used to describe experimental data by fitting appropriately the effective mass and the velocity matrix element $\mathbf{v}_{cv}(\mathbf{\Gamma})$.\cite{Komkov09} For the examples shown here we use the values
\begin{equation}
	m^{\ast}_{cv} = 0.0553 m_o \, \text{ and } \, \hbar v_{cv}^{x}(\boldsymbol \Gamma) = 10.3 \text{ eV \AA}, \notag
\end{equation}
which correspond to the pair formed by one of the lowest conduction bands and one of the heavy hole bands; $m_o$ is the electron mass.\\ 


\subsection{The 14-band $\mathbf{k} \cdot \mathbf{p}$ model \label{S:14bandModel}}

\begin{figure}
\centering
\includegraphics[width=6.5cm]{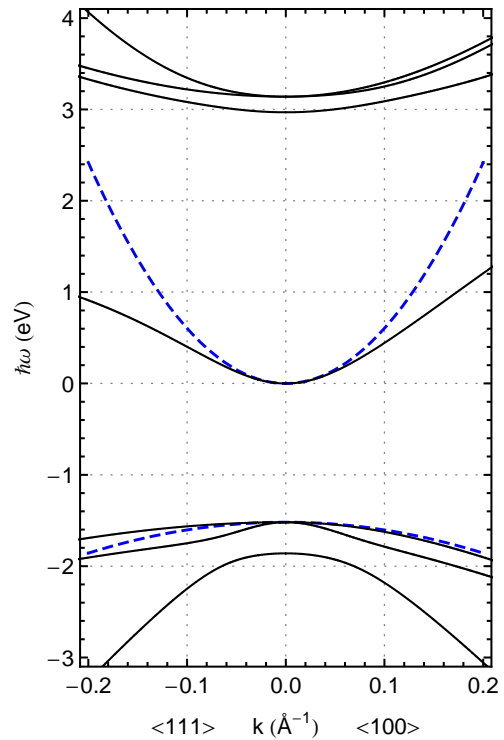}
\caption{(Color online) Band structure calculated with the 14-band model for GaAs (solid lines). For reference, the parabolic bands used for the two-band model are also shown (dashed blue lines); these correspond to the lowest conduction and the heavy hole bands. Notice the significant nonparabolicity of the bands, for example the band warping of the heavy and light holes. Spin splitting of the bands ($\lesssim 1\text{ meV}$) is not shown.}
\label{F:BandStructure}
\end{figure}

The 14-band model considered here consists of six p-like valence bands, two s-like conduction bands, and six p-like conduction bands; these are shown in Fig.~\ref{F:BandStructure}. Effects beyond this set of bands are included using L\"owdin perturbation theory.\cite{Lowdin51} The relevant aspects of the model were summarized by Wahlstrand and Sipe; \cite{Wahlstrand10} for further details we refer the reader to the work of Bhat and Sipe,\cite{Bhat06} whose numerical values for the relevant parameters are the ones used here.\\


For the $\mathbf{k} \cdot \mathbf{p}$ model calculation we use the basis of the $\Gamma$ point\cite{LewYanVoon09} to write
\begin{equation}
	u_{n \mathbf{k}}(\mathbf{x}) = \sum_{n'} u_{n' \mathbf{\Gamma}}(\mathbf{x}) \mathcal{C}_{n' n} (\mathbf{k}), \notag
\end{equation}
where the coefficients $\mathcal{C}_{n' n} (\mathbf{k})$ can be arranged in a unitary matrix $\mathcal{C} (\mathbf{k})$. Once these coefficients are found, we can calculate all the required matrix elements. The $u_{n \mathbf{k}}(\mathbf{x})$ satisfy
\begin{equation}
	H_{\mathbf{k}} u_{n \mathbf{k}}(\mathbf{x}) = \hbar \omega_{n}(\mathbf{k}) u_{n \mathbf{k}}(\mathbf{x}), \notag
\end{equation}
where
\begin{equation}
	H_{\mathbf{k}} \equiv e^{-i \mathbf{k} \cdot \mathbf{x}} \mathcal{H}_{o} e^{i \mathbf{k} \cdot \mathbf{x}}. \notag
\end{equation}
Thus, the elements of $\mathcal{C} (\mathbf{k})$ satisfy
\begin{equation}
	\sum_{n_1'} \left( H_{\mathbf{k}} \right)_{n_1 n_1'} \mathcal{C}_{n_1' n_2}(\mathbf{k}) = \hbar \omega_{n_2}(\mathbf{k}) \mathcal{C}_{n_1 n_2}(\mathbf{k}), \label{E:C_matrix_diagonalization_all_k}
\end{equation}
and they can be found by simple diagonalization of the $\mathbf{k} \cdot \mathbf{p}$ Hamiltonian,\cite{Wahlstrand10}
\begin{equation}
	\left( H_{\mathbf{k}} \right)_{n_1 n_2} \equiv \frac{1}{\Omega_{\text{cell}}} \int_{\text{cell}} u_{n_1 \mathbf{\Gamma}}^{\ast}(\mathbf{x}) H_{\mathbf{k}} u_{n_2 \mathbf{\Gamma}}(\mathbf{x}) \, d \mathbf{x}, \notag
\end{equation}
at each $\mathbf{k}$.\\

The dc field couples Bloch states along lines parallel to it. Consequently, the phase of the Bloch states along these lines needs to be smooth in order to calculate the matrix elements in Eq.~\eqref{E:DefLaxConnection}. In Sec.~\ref{S:THEORETICAL_FRAMEWORK} we implicitly assumed that it was possible to find Bloch states conforming to such requirement. Numerical diagonalization of the $\mathbf{k} \cdot \mathbf{p}$ Hamiltonian introduces a random phase, which is incompatible with this assumption; however, there can still be a random phase between different lines parallel to the dc field. Therefore, we diagonalize on a plane perpendicular to the dc field that crosses the $\Gamma$ point ($k_{\parallel}=0$) to determine $\mathcal{C}(\mathbf{k}_{\perp})$.\\

In order to proceed along lines parallel to the dc field we follow the method used by Gharamani and Sipe, which can deal with points and lines of degeneracy.\cite{Gharamani93} This method also allows us to calculate the velocity matrix elements $\mathbf{V}_{n_1 n_2}(\mathbf{k})$ without finding first the Berry-type phases in Eq.~\eqref{E:Berry_typePhase}. We define
\begin{equation}
	C(\mathbf{k}) \equiv \mathcal{C}(\mathbf{k}) e^{-i \Xi(\mathbf{k})}, \label{E:ModifiedKohnLuttingerCoefficients}
\end{equation}
which satisfies $C(\mathbf{k}_{\perp})=\mathcal{C}(\mathbf{k}_{\perp})$. We can write the velocity matrix elements in terms of these matrices as\cite{Wahlstrand10, Bhat06}
\begin{equation}
	\mathbf{V}_{n_1 n_2}(\mathbf{k}) = \frac{1}{\hbar} \sum_{n_1', n_2'} C_{n_1 n_1'}^{\dagger}(\mathbf{k}) (\nabla_{\mathbf{k}} H_{\mathbf{k}})_{n_1' n_2'} C_{n_2' n_2}(\mathbf{k}). \label{E:VelocityKL}
\end{equation}
For $k_{\parallel} \ne 0$ we require\cite{Wahlstrand10, Gharamani93}
\begin{equation}
	ie \frac{\partial}{\partial k_{\parallel}} C_{n_1 n_2}(\mathbf{k}) - \sum_{n_2'} C_{n_1 n_2'}(\mathbf{k}) \mu_{n_2' n_2}^{\text{dc}}(\mathbf{k}) = 0, \notag
\end{equation}
using Eq.~\eqref{E:DefDipoleMatrixElement} and Eq.~\eqref{E:VelocityKL}. Recall that $\mu^{\text{dc}}(\mathbf{k})$ denotes the \emph{full} matrix, which includes the coupling between conduction and valence bands.\\

Once the matrix $\mathbf{V}(\mathbf{k})$ has been calculated over the region of interest in the Brillouin zone, we can solve numerically Eq.~\eqref{E:dcInterCouplingDiffEq} to find the interband coupling matrix $\tilde{M}(\mathbf{k})$ and the transformed dipole moment matrix defined in Eq.~\eqref{E:TransformedDipole}. Note that the $\tilde{M}_{n_1 n_2}(\mathbf{k})$ defined here are equal to the matrix elements $\tilde{m}_{n_1 n_2}(\mathbf{k}_{\perp};k_{\parallel}/\epsilon)$ used by Wahlstrand and Sipe\cite{Wahlstrand10} except for a phase,
\begin{equation}
	\tilde{M}_{n_1 n_2}(\mathbf{k}) = \tilde{m}_{n_1 n_2}(\mathbf{k}_{\perp};k_{\parallel}/\epsilon) e^{-i\int_{0}^{k_{\parallel}} \omega_{n_1 n_2}(\mathbf{k}_{\perp} + k_{\parallel}' \hat{\mathbf{e}}_{\text{dc}}) dk_{\parallel}' /\epsilon}, \notag
\end{equation}
where we introduce the reduced electric field
\begin{equation}
	\epsilon \equiv \frac{e E^{\text{dc}}}{\hbar}. \label{E:ReducedElectricField}
\end{equation}\\

Finally, we need to find the transformed overlap matrix $K(\mathbf{k}_1, \mathbf{k}_2)$ defined in Eq.~\eqref{E:OverlapKP}. We write 
\begin{equation}
	e^{i \Xi(\mathbf{k}_1)} \Delta(\mathbf{k}_1, \mathbf{k}_2) e^{-i \Xi(\mathbf{k}_2)} = C^{\dagger}(\mathbf{k}_1) C(\mathbf{k}_2) \notag
\end{equation}
in terms of $C(\mathbf{k})$. In this form, $K(\mathbf{k}_1, \mathbf{k}_2)$ can be found from
\begin{equation}
	K(\mathbf{k}_1, \mathbf{k}_2) = \tilde{M}^{\dagger}(\mathbf{k}_1) C^{\dagger}(\mathbf{k}_1) C(\mathbf{k}_2) \tilde{M}(\mathbf{k}_2). \notag
\end{equation}
In the next subsection it will be convenient to work in terms of the matrix
\begin{equation}
	R(\mathbf{k}) \equiv C(\mathbf{k}) \tilde{M}(\mathbf{k}), \notag
\end{equation}
so that 
\begin{equation}
	K(\mathbf{k}_1, \mathbf{k}_2) = R^{\dagger}(\mathbf{k}_1) R(\mathbf{k}_2). \notag
\end{equation}\\
   

\subsection{Time evolution of the interband polarization \label{S:Time_ev_InterPol}}

The interband polarization $\bar{\mathcal{P}}_{cv}^{(i)}(\mathbf{k},t)$ in Eq.~\eqref{E:InterbandPolRWAEq} oscillates due to the term  with the energy difference $\hbar \omega_{cv}(\mathbf{k})$, and moves in \textbf{k}-space due to the gradient term. For these reasons we introduce a \emph{slowly varying} interband polarization, $\bar{\mathfrak{P}}^{(i)}_{cv}(\boldsymbol \kappa,t)$, in the \emph{moving frame}, so that
\begin{equation}
	\bar{\mathcal{P}}_{cv}^{(i)}(\mathbf{k},t) = \bar{\mathfrak{P}}^{(i)}_{cv}(\boldsymbol \kappa,t) e^{- i \int_{0}^{t}(\omega_{cv}(\boldsymbol \kappa + \epsilon t' \hat{\mathbf{e}}^{\text{dc}}) - \omega_o) \, dt'}, \label{E:EnvelopeMovInterPol}
\end{equation}
where $\boldsymbol \kappa \equiv \mathbf{k} - \epsilon t \hat{\mathbf{e}}^{\text{dc}}$ is a \emph{moving} coordinate in reciprocal space, and $\epsilon$ is the reduced electric field, Eq.~\eqref{E:ReducedElectricField}. The new interband polarization satisfies the dynamical equation
\begin{equation}
	i \hbar \frac{\partial}{\partial t} \bar{\mathfrak{P}}^{(i)}_{cv}(\boldsymbol \kappa,t) - \mathfrak{S}_{cv}^{(i)}(\boldsymbol \kappa,t) = - \mathcal{E}^{\text{opt}}_{i} (t) \bar{\mathfrak{T}}_{cv}^{i}(\boldsymbol \kappa,t), \label{E:EvolutionEnvelopeMovInterPol}
\end{equation}
where, similarly to Eq.~\eqref{E:EnvelopeMovInterPol}, we define
\begin{equation}
	\mathfrak{S}^{(i)}_{cv}(\boldsymbol \kappa,t) \equiv \Sigma^{\bar{\mathcal{P}}^{(i)}}_{cv}(\boldsymbol \kappa + \epsilon t \hat{\mathbf{e}}^{\text{dc}}, t) e^{i \int_{0}^{t}(\omega_{cv}(\boldsymbol \kappa + \epsilon t' \hat{\mathbf{e}}^{\text{dc}}) - \omega_o) \, dt'} \label{E:EnvelopeMovCoul}
\end{equation} 
and
\begin{equation}
	\bar{\mathfrak{T}}^{i}_{cv}(\boldsymbol \kappa,t) \equiv \bar{\Theta}^{i}_{cv} (\boldsymbol \kappa + \epsilon t \hat{\mathbf{e}}^{\text{dc}}) e^{i \int_{0}^{t}(\omega_{cv}(\boldsymbol \kappa + \epsilon t' \hat{\mathbf{e}}^{\text{dc}}) - \omega_o) \, dt'}. \label{E:EnvelopeMovDipole}
\end{equation}\\

We solve Eq.~\eqref{E:EvolutionEnvelopeMovInterPol} numerically using a 4th order Runge-Kutta algorithm.\cite{Press07} Once $\bar{\mathfrak{P}}_{cv}^{(i)}(\boldsymbol \kappa,t)$ has been found for a given time step, the Coulomb term, $\bar{\mathfrak{S}}^{(i)}_{cv}(\boldsymbol \kappa,t)$, is calculated in three steps. First we use Eq.~\eqref{E:EnvelopeMovInterPol} to go back to the lab frame, including the fast oscillations, and transform $\bar{\mathcal{P}}_{c v}^{(i)}(\mathbf{k},t)$ using the matrix $R(\mathbf{k})$ (see Sec.~\ref{S:14bandModel}). Next, we calculate the convolution of the resulting matrix with the Coulomb potential,
\begin{multline}
	\mathfrak{C}^{(i)}_{n_1 n_2}(\mathbf{k},t) \equiv \sum_{c' v'}  \int \frac{d \mathbf{q}}{(2 \pi)^3} U(\mathbf{q}) \\
	\times R_{n_1 c'}(\mathbf{k} - \mathbf{q}) \bar{\mathcal{P}}_{c' v'}^{(i)}(\mathbf{k}-\mathbf{q},t) R_{v' n_2}^{\dagger}(\mathbf{k} - \mathbf{q}), \label{E:FirstStepCoulomb}
\end{multline} 
using a fast Fourier transform implementation.\cite{Press07, Frigo05} Finally, we invert the matrix transform and remove the fast oscillations using the phase factor in Eq.~\eqref{E:EnvelopeMovCoul} to identify a useful expression for $\mathfrak{S}^{(i)}_{cv}(\boldsymbol \kappa,t)$, 
\begin{multline}
	\mathfrak{S}^{(i)}_{cv}(\boldsymbol \kappa,t) = e^{i \int_{0}^{t}(\omega_{cv}(\boldsymbol \kappa + \epsilon t' \hat{\mathbf{e}}^{\text{dc}}) - \omega_o) \, dt'}\\
	\times \sum_{n_1' n_2'} \bigg[ R_{c n_1'}^{\dagger}(\boldsymbol \kappa + \epsilon t \hat{\mathbf{e}}^{\text{dc}}) \mathfrak{C}^{(i)}_{n_1' n_2'}(\boldsymbol \kappa + \epsilon t \hat{\mathbf{e}}^{\text{dc}},t) R_{n_2' v}(\boldsymbol \kappa + \epsilon t \hat{\mathbf{e}}^{\text{dc}}) \notag \\
	+ R_{c n_1'}^{\dagger}(\boldsymbol \kappa + \epsilon t \hat{\mathbf{e}}^{\text{dc}}) \mathfrak{C}^{(i) \dagger}_{n_1' n_2'}(\boldsymbol \kappa + \epsilon t \hat{\mathbf{e}}^{\text{dc}},t) R_{n_2' v}(\boldsymbol \kappa + \epsilon t \hat{\mathbf{e}}^{\text{dc}}) \bigg].
\end{multline}
Note that the Coulomb term $\bar{\mathfrak{S}}^{(i)}_{cv}(\boldsymbol \kappa,t)$ couples the interband polarization for all possible pairs of conduction and valence bands (see Eq.~\eqref{E:FirstStepCoulomb}). In order to reduce the number of pairs involved in the numerical calculation for the 14-band model, we neglect the contribution from the six upper conduction bands, which will oscillate much faster due to their energy separation with respect to the valence bands. The Coulomb potential used in the calculations is assumed to have the form
\begin{equation}
	U(\mathbf{q}) = \frac{4 \pi e^2}{\epsilon_{\text{b}}} \frac{1}{q^2}, \label{E:BareCoulomb} 
\end{equation}
where $\epsilon_{\text{b}}$ is the background dielectric constant; we use the value $\epsilon_{\text{b}}=12.9$ for GaAs.\footnote{The divergence of the bare Coulomb potential at $\mathbf{q}=0$ is avoided with a simple lower cut-off at the value of $U(\sqrt{3 (\delta \kappa_{\perp})^2})$, where $\delta \kappa_{\perp}$ is step used in the rectangular grid for $\boldsymbol \kappa_{\perp}$. The error introduced by this cut-off was reduced by decreasing $\delta \kappa_{\perp}$ until no significant modification in the absorption spectrum was found.} \\

For the time propagation of $\bar{\mathfrak{P}}_{cv}^{(i)}(\boldsymbol \kappa,t)$ in the moving frame, we perform our calculation within a window given by $|\kappa_{\parallel}|<0.2 \, \text{\AA}^{-1}$ and $|\kappa_{\perp}|<0.1 \, \text{\AA}^{-1}$, except for some of the runs involving two bands where we find we need a larger range, $|\kappa_{\perp}|<0.2 \, \text{\AA}^{-1}$, to ensure convergence . We let the system evolve for enough time so that the interband polarization in the lab frame is inside the window $-0.25 \, \text{\AA}^{-1}<k_{\parallel}<0.5 \, \text{\AA}^{-1}$. Previous calculations have shown that this range is appropriate for calculations of the optical absorption.\cite{Wahlstrand10, Hader97} The numerical calculations are very demanding when the Coulomb term is included. For example, in the 14 band model using 44 AMD Opteron$\texttrademark$ 6176 SE processors in parallel, each time step takes about $1100 \text{ s}$ for $E^{\text{dc}}=66 \text{ kV/cm}$. For lower dc fields the calculation is even more time consuming, since the required step sizes in time and $\kappa_{\parallel}$ are smaller.\cite{Wahlstrand10} \\

\begin{figure}
\centering
\includegraphics[width=8.5cm]{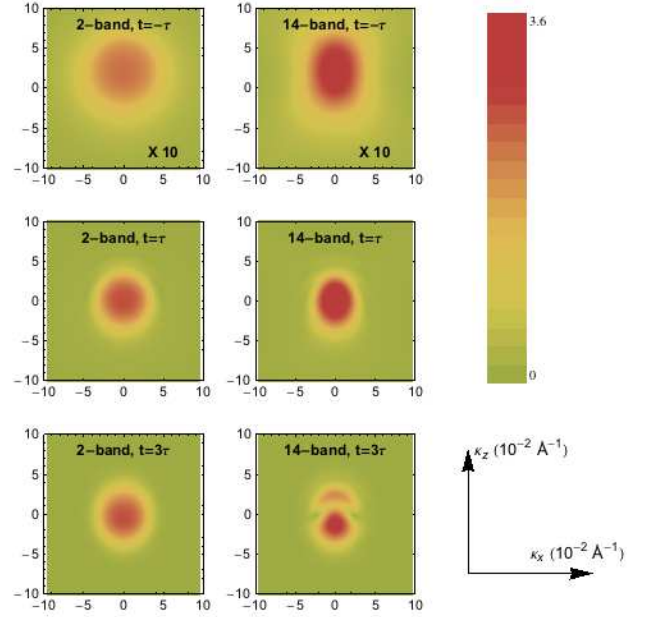}
\caption{(Color online) Snapshots of $| \mathfrak{N}(\boldsymbol \kappa, t) |$ in the independent particle approximation for the two-band (left column) and 14-band (right column) models in arbitrary units (the plots for $t=-\tau$ are amplified 10 times). The dc field is oriented along the [001] direction and its reduced value is $\epsilon=1/(\text{ps \AA})$, which corresponds to 66 kV/cm. The horizontal and vertical axes represent $\kappa_x=k_x$ and  $\kappa_z$, respectively. The optical pulse has a duration of $\tau=16 \text{ fs}$ and it is oriented along the [100] direction; for simplicity, we tune its frequency to the band gap of GaAs, $\hbar \omega_o = 1.519 \text{ eV}$.}
\label{F:AbsInteg__eps1__2bandvs14band}
\end{figure} 

\begin{figure}
\centering
\includegraphics[width=8.5cm]{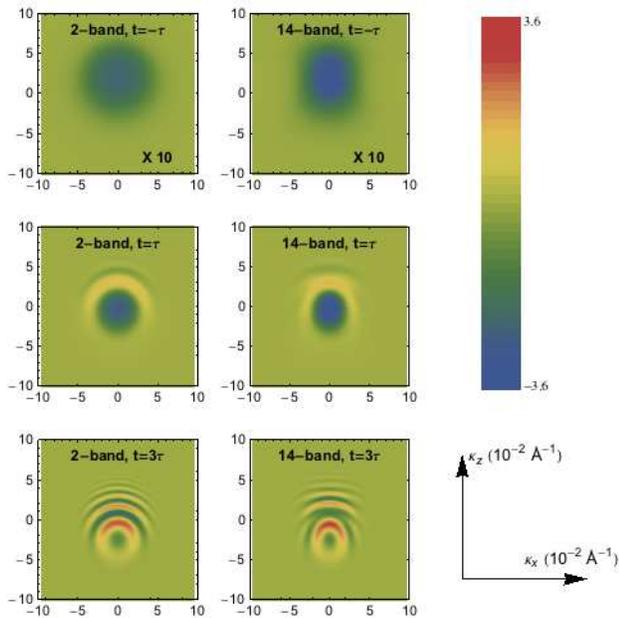}
\caption{(Color online) Snapshots of $\text{Im} \left[ \mathfrak{N}(\boldsymbol \kappa, t) \right]$ for the same parameters used in Fig.~\ref{F:AbsInteg__eps1__2bandvs14band}.}
\label{F:ImInteg__eps1__2bandvs14band}
\end{figure}

We illustrate the time evolution of the interband polarization with the function
\begin{equation}
	\mathcal{N}^{(i)}(\mathbf{k}, t) = \sum_{c v} \bar{\mathcal{P}}_{c v}^{(i)}(\mathbf{k}, t) \bar{\Theta}^{i}_{v c}(\mathbf{k}). \label{E:AbsorpIntLab}
\end{equation}
The imaginary part of this function is proportional to the density of excited carriers after multiplication by the optical field envelope and integration over $\mathbf{k}$ and time. In the moving frame Eq.~\eqref{E:AbsorpIntLab} becomes
\begin{multline} 
	\mathfrak{N}^{(i)}(\boldsymbol \kappa, t) = \sum_{c v} \bar{\mathfrak{P}}_{c v}^{(i)}(\boldsymbol \kappa,t) e^{-i \int_{0}^{t}(\omega_{cv}(\boldsymbol \kappa + \epsilon t' \hat{\mathbf{e}}^{\text{dc}}) - \omega_o) \, dt' }\\
	 \times \bar{\Theta}^{i}_{vc}(\boldsymbol \kappa + \epsilon t \hat{\mathbf{e}}^{\text{dc}}), \label{E:AbsorpIntMov}
\end{multline}
where $\mathfrak{N}^{(i)}(\boldsymbol \kappa, t) \equiv \mathcal{N}^{(i)}(\mathbf{k}, t)$. The color maps in Figs.~\ref{F:AbsInteg__eps1__2bandvs14band} and \ref{F:ImInteg__eps1__2bandvs14band} show the absolute value and imaginary part of $\mathfrak{N}^{(i)}(\boldsymbol \kappa, t)$ for the two models considered here. We choose the [001] direction for the dc field, $\hat{\mathbf{e}}^{\text{dc}}=\hat{\mathbf{z}}$, and a positive reduced electric field, $\epsilon>0$; therefore, the region of non-zero $\mathcal{N}^{(i)}(\mathbf{k}, t) $ moves in the positive $k_z$ direction in the lab frame.  For simplicity, we first calculate $\mathfrak{N}^{(i)}(\boldsymbol \kappa, t)$  neglecting the Coulomb interaction; later we will show that the effect of the Coulomb term is mainly an enhancement of the interband polarization (see Sec.~\ref{S:Results}).\\

In order to explain the shape and motion of $\mathfrak{N}^{(i)}(\boldsymbol \kappa, t)$ we consider first the contribution from $\bar{\mathfrak{P}}_{c v}^{(i)}(\boldsymbol \kappa,t)$ to Eq.~\eqref{E:AbsorpIntMov}. In the independent particle approximation the solution to Eq.~\eqref{E:EvolutionEnvelopeMovInterPol} can be written as
\begin{equation}
	\bar{\mathfrak{P}}_{cv}^{(i)}(\boldsymbol \kappa,t) = \frac{i}{\hbar} \int_{t_o}^{t} \mathcal{E}^{\text{opt}}_{i}(t') \bar{\mathfrak{T}}^{i}_{cv}(\boldsymbol \kappa,t') \, dt'. \label{E:InterbandPolarizationMovingFrame}
\end{equation}
In the integrand of this expression, the optical field envelope is centered at $t'=0$ but the term with the transformed dipole moment matrix, $\bar{\mathfrak{T}}^{i}_{cv}(\boldsymbol \kappa,t')$, is centered at a value of $t'$ that depends on $\kappa_z$ (see Eq.~\eqref{E:EnvelopeMovDipole}). Therefore, when we integrate over $t'$, the distribution of $\bar{\mathfrak{P}}_{cv}^{(i)}(\boldsymbol \kappa,t)$ is not centered at $\kappa=0$ for all times, as can be seen in Fig.~\ref{F:AbsInteg__eps1__2bandvs14band}; there is a visible shift, in the positive $\kappa_z$ direction, of the center of the distribution for times $t<0$. The reason for this shift is that $\bar{\mathfrak{T}}^{(i)}_{cv}(\boldsymbol \kappa,t')$ contributes more significantly for $\kappa_z>0$ when the upper limit of the integral in Eq.~\eqref{E:InterbandPolarizationMovingFrame} is negative; only when $t$ is positive and sufficiently large, is there a balance between the contributions from positive and negative $\kappa_z$, and the center of the distribution of $\bar{\mathfrak{P}}_{cv}^{(i)}(\boldsymbol \kappa,t)$ remains essentially at $\kappa_z=0$.\\

The phase factor in Eq.~\eqref{E:AbsorpIntMov} contributes oscillations to $\mathfrak{N}^{(i)}(\boldsymbol \kappa, t)$, which are noticeable in Fig.~\ref{F:ImInteg__eps1__2bandvs14band}. The frequency of the oscillation increases with time as expected for accelerated carriers, but the center of the oscillations, where they are slower, moves in time and does not coincide with the center of the distribution of $|\bar{\mathfrak{P}}_{cv}^{(i)}(\boldsymbol \kappa,t)|$ for all times; furthermore, the center of $\bar{\Theta}^{i}_{vc}(\boldsymbol \kappa + \epsilon t \hat{\mathbf{e}}^{\text{dc}})$ in Eq.~\eqref{E:AbsorpIntMov} also moves in time due to the shift $\epsilon t \hat{\mathbf{e}}^{\text{dc}}$. As a result, the oscillations of $\mathfrak{N}^{(i)}(\boldsymbol \kappa, t)$ for times $t>0$ are centered at $\kappa_z<0$ (for instance, see $t=3 \tau$ in Fig.~\ref{F:ImInteg__eps1__2bandvs14band}). The motion of the center of the oscillations can be illustrated for low dc fields such that the approximation
\begin{equation}
	\int_{0}^{t}(\omega_{cv}(\boldsymbol \kappa + \epsilon t' \hat{\mathbf{e}}^{\text{dc}}) - \omega_o) \, dt' \approx \left( \omega_{cv}(\boldsymbol \kappa + \epsilon t \hat{\mathbf{e}}^{\text{dc}}) - \omega_o \right) t \label{E:ApproxPhase}
\end{equation}
is valid. In this case it is clear that the center of the oscillations will move in the $\kappa_z$ direction at a rate given by $\epsilon$. Although in our calculations $\epsilon$ is not sufficiently small for Eq.~\eqref{E:ApproxPhase} to be valid, Fig.~\ref{F:ImInteg__eps1__2bandvs14band} illustrates that the center of the oscillations still moves in the $\kappa_z$ direction.\\ 

Qualitatively the function $\mathfrak{N}^{(i)}(\boldsymbol \kappa, t)$ behaves similarly in the two-band and 14-band models. As expected, the round shapes in the two-band model are modified in the 14-band model due to the non-parabolicity of the band energies and the more complicated structure of the transformed dipole moment matrix. The difference between the two models is more visible in the absolute value of $\mathfrak{N}^{(i)}(\boldsymbol \kappa, t)$, shown in Fig.~\ref{F:AbsInteg__eps1__2bandvs14band}, where the deviations from the simple two-band case are particularly visible at times when the pulse has stopped driving the system and carriers are simply accelerated by the dc field (see $t=3\tau$ in Fig.~\ref{F:AbsInteg__eps1__2bandvs14band}).\\

Once the interband polarization, Eq.~\eqref{E:EnvelopeMovInterPol}, is calculated in time-domain, we proceed to calculate the absorption coefficient $\alpha(\omega)$ as described in Sec.~\ref{S:AbsorptionCoefficient}. In the next section we discuss the results for the absorption coefficient with and without including the Coulomb interaction, for different dc fields as calculated using the band models just described.\\ 


\begin{figure*}
\centering
\includegraphics[width=11.0cm]{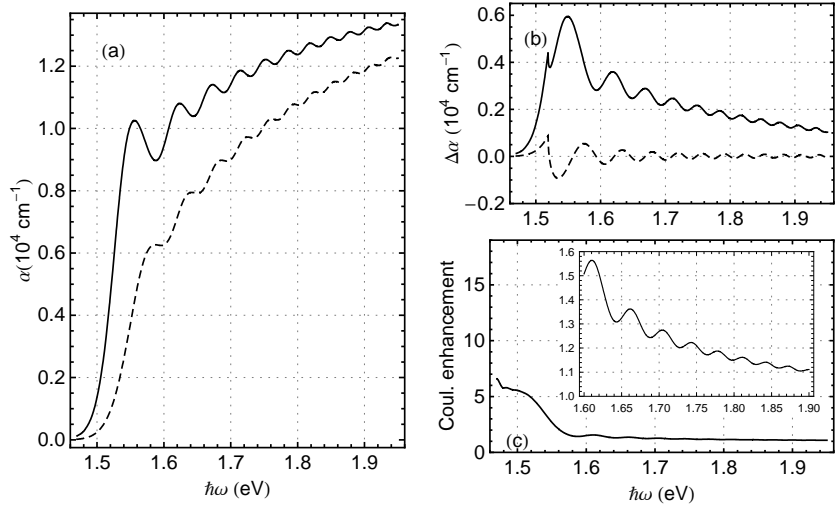}
\caption{Franz-Keldysh spectrum for a 66 kV/cm dc field calculated in the two-band model. (a) Absorption coefficient with Coulomb interaction (solid line) and without (dashed line). (b) Difference with respect to the zero-field absorption in the independent particle approximation for $\alpha(\omega)$ calculated with Coulomb interaction (solid line) and without (dashed line). (c) Ratio between the absorption coefficient with and without Coulomb interaction; the inset shows the energy range from 1.6 to 1.9 eV in more detail.}
\label{F:eps1__2band__Ind_Coul}
\end{figure*} 


\section{RESULTS \label{S:Results}}

In this section we study the absorption spectra calculated from the two models discussed in Sec.~\ref{S:Calculations}, and examine the effect of the Coulomb interaction. The first set of results corresponds to $E^{\text{dc}}=66 \text{ kV/cm}$ ($\epsilon=1/(\text{ps \AA})$), which is the most extensively discussed field in the work of Wahlstrand and Sipe.\cite{Wahlstrand10} Here we show how the excitonic effects modify their results. We also consider a lower dc field, $E^{\text{dc}}=44 \text{ kV/cm}$ ($\epsilon=2/(3 \text{ ps \AA})$), where the Coulomb interaction modifies the absorption spectrum more significantly because the force due to $E^{\text{dc}}$, which pulls apart the electron-hole pair, is weaker.\\ 


\begin{figure*}
\centering
\includegraphics[width=11.0cm]{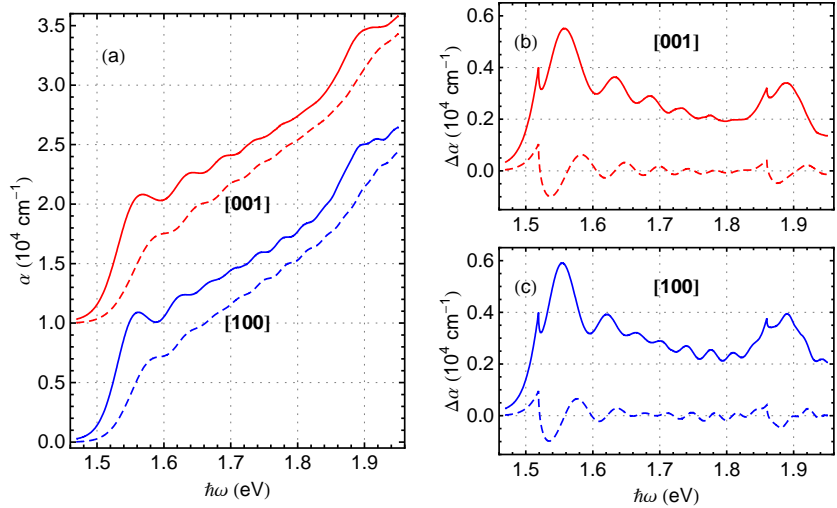}
\caption{(Color online) Franz-Keldysh spectrum for a 66 kV/cm dc field pointing along [001] calculated in the 14-band model. (a) Absorption coefficient with Coulomb interaction (solid lines) and without (dashed lines) for light polarized along [001] (red lines) and [100] (blue lines). The spectra are offset vertically for clarity. [(b) and (c)] Difference with respect to the zero-field absorption in the independent particle approximation for $\alpha(\omega)$ calculated with Coulomb interaction (solid lines) and without (dashed lines), for light polarized along [001] and [100].}
\label{F:eps1__x_z__Ind_Coul}
\end{figure*} 

The absorption coefficient for $E^{\text{dc}}=66 \text{ kV/cm}$ in the parabolic two-band model is shown in Fig.~\ref{F:eps1__2band__Ind_Coul}a. Note the absorption tail below the band gap and oscillations above it, the typical features of the Franz-Keldysh effect. In order to examine how the dc field modifies the zero-field absorption, we plot the difference between the absorption coefficient (with and without Coulomb interaction) and the zero-field absorption coefficient in the independent particle approximation (see Fig.~\ref{F:eps1__2band__Ind_Coul}b). Without Coulomb interaction this difference is simply the differential electroabsorption, and our numerical calculations agree with the Airy function result,\cite{HaugKoch04} 
\begin{equation}
	\Delta \alpha (\omega) \propto \frac{1}{\pi} \{ \left[ \text{Ai}^{\prime}(x) \right]^2 - x \text{Ai}^2(x) - (-x)^{1/2} \theta(-x) \}, \notag
\end{equation}
where $x=(E_g-\hbar \omega)/(\hbar \Omega_{cv})$ is the energy relative to the band gap scaled by the electro-optic function $\Omega_{cv} \equiv (\hbar \epsilon^2 / 2 m_{cv}^{\ast})^2$ and $\theta(x)$ is the Heaviside step function. The Coulomb interaction enhances the absorption coefficient, especially near the bandgap, as can be seen in the ratio between the Franz-Keldysh spectra with and without electron-hole interaction plotted in Fig.~\ref{F:eps1__2band__Ind_Coul}c; at higher energies the enhancement factor becomes flatter. Our results are slightly different from previous two-band model calculations with Coulomb interaction that consider the contributions from the hydrogenic exciton wavefunctions only at the origin.\cite{Blossey70, Dow71, Fauchier74} In our scheme this would require fixing the dipole moment matrix element to the value at the $\Gamma$ point,\cite{HaugKoch04} 
\begin{equation}
	\boldsymbol \mu_{cv}(\mathbf{k}) = \frac{e \mathbf{v}_{cv}(\boldsymbol \Gamma)}{i \omega_{cv}(\boldsymbol \Gamma)}, \label{E:FixedDipole}
\end{equation}
instead of fixing only the velocity matrix element (see Eq.~\eqref{E:VelMatElTwoBand}). The numerical calculations assuming Eq.~\eqref{E:FixedDipole} are more demanding in our scheme because the dipole moment matrix element does not decrease away from the $\Gamma$ point. The absorption coefficient in both cases is essentially the same up to a numerical factor, the enhancement being larger for Eq.~\eqref{E:FixedDipole}.\\


\begin{figure}
\centering
\includegraphics[width=8.5cm]{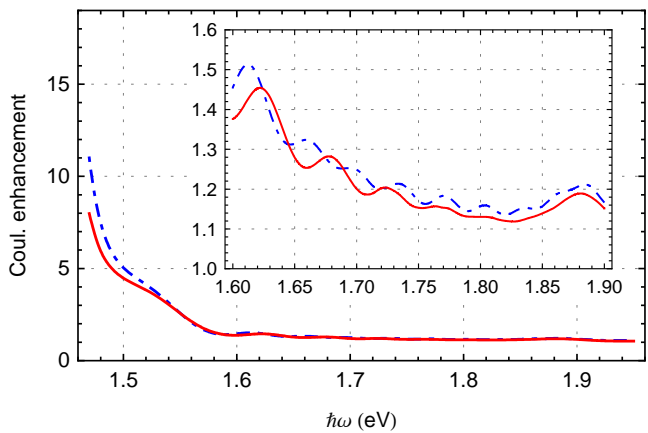}
\caption{(Color Online) Ratio between the absorption coefficient with and without Coulomb interaction calculated in the 14-band model as in Fig~\ref{F:eps1__x_z__Ind_Coul}. Two light polarizations are shown, along [100] (dash-dotted blue line) and  [001] (solid red line). The inset shows the energy range from $1.6$ to $1.9$ eV in more detail.}
\label{F:CoulEnhancement__eps1__x_z}
\end{figure} 

\begin{figure}
\centering
\includegraphics[width=8.5cm]{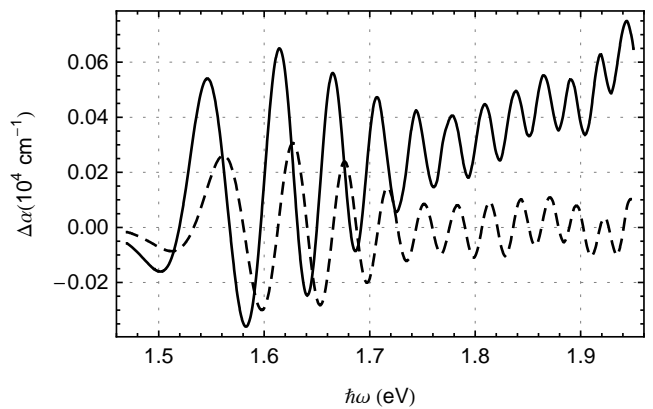}
\caption{Difference between the absorption coefficients for [100] and [001] polarized light in Fig.~\ref{F:eps1__x_z__Ind_Coul}, with Coulomb interaction (solid line) and without (dashed line).}
\label{F:OptPolDiff__eps1__x_z}
\end{figure}

In the 14-band model the absorption coefficient displays additional features as shown in Fig.~\ref{F:eps1__x_z__Ind_Coul}a for [001] and [100] polarized light; in both cases the dc field is oriented along [001]. For example, there is a dependence on the polarization of the optical field, which can be seen more easily in the difference with respect to the zero-field absorption (compare Figs.~\ref{F:eps1__x_z__Ind_Coul}b and c). Not surprisingly, the Coulomb term, which is isotropic, enhances the absorption coefficient for both polarizations in approximately the same way (see Fig.~\ref{F:CoulEnhancement__eps1__x_z}), favoring slightly the transverse configuration ($\mathbf{E}^{\text{dc}} \perp \mathbf{E}^{\text{opt}}$) below the bandgap. A more detailed comparison is shown in Fig.~\ref{F:OptPolDiff__eps1__x_z}; in addition to the enhancement, a phase shift and a vertical offset in the difference between the two polarizations arises when the Coulomb interaction is included. Finally, another feature not captured in the two-band model is the `bump' in $\alpha(\omega)$ around 1.86 eV, which is due to the contribution of the split-off bands; the Coulomb interaction significantly enhances this contribution.\\


\begin{figure}
\centering
\includegraphics[width=8.5cm]{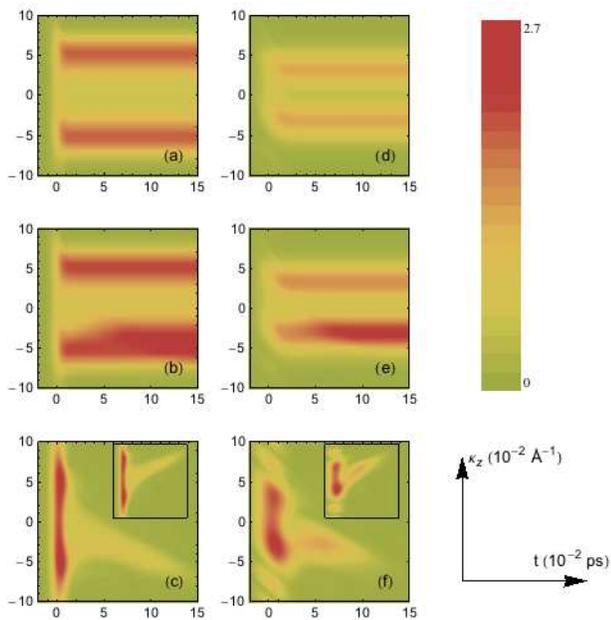}
\caption{(Color online) Time evolution of $|\bar{\mathfrak{P}}_{cv}(\boldsymbol \kappa,t)|$ in the independent particle approximation (panels (a) and (d)) and including Coulomb interaction (panels (b) and (e)) in arbitrary units for a 66 kV/cm dc field oriented along the [001] direction and light polarized along the [100] direction, as calculated using the two-band and 14-band models. The horizontal and vertical axes represent time and wavevector along the z-direction in the moving frame, respectively; the coordinates perpendicular to the dc field are fixed at $\kappa_x = 0$ and $\kappa_y = 0$. We also include plots of $|\mathfrak{S}_{cv}(\boldsymbol \kappa,t)|$ (panels (c) and (f), in arbitrary units) in the moving frame; the insets in these plots show $|\Sigma^{\bar{\mathcal{P}}}_{cv}(\mathbf{k},t)|$ in the lab frame, where $\kappa_z$ is replaced by $k_z$ for the vertical axis. [(a)-(c)] Two-band model with a pulse duration of $\tau=8 \text{ fs}$ and center frequency such that $\hbar \omega_o =1.719 \text{ eV}$.  [(d)-(f)] 14-band model with a pulse duration of $\tau=16 \text{ fs}$ and center frequency such that $\hbar \omega_o =1.619 \text{ eV}$; the two bands chosen in this case where the lowest conduction band and the highest valence band.}
\label{F:G1_CoulTerm__eps1__2bandvs14band}
\end{figure}

The enhancement of the absorption coefficient due to the Coulomb interaction can be traced back to the interband polarization. We illustrate this point in Fig.~\ref{F:G1_CoulTerm__eps1__2bandvs14band}, where $|\bar{\mathfrak{P}}_{cv}(\boldsymbol \kappa,t)|$ and $|\bar{\mathfrak{S}}_{cv}(\boldsymbol \kappa,t)|$ are plotted in the moving frame for fixed $\boldsymbol \kappa_{\perp} = 0 \mathbf{\hat{x}} + 0 \mathbf{\hat{y}}$ as they evolve in time. Note that $|\bar{\mathfrak{P}}_{cv}(\boldsymbol \kappa,t)|$ is peaked at values of $\kappa_z$ such that the energy difference between the bands $c$ and $v$ is tuned to the energy associated with the the center frequency of the pulse, $\hbar \omega_o$. Since we have not included dephasing, it is also clear that $|\bar{\mathfrak{P}}_{cv}(\boldsymbol \kappa,t)|$ remains essentially unchanged after the pulse has stopped driving the system. The behavior of $|\bar{\mathfrak{S}}_{cv}(\boldsymbol \kappa,t)|$ is quite different; although it has the same two peaks present in the interband polarization, the Coulomb term decreases significantly and eventually vanishes when the pulse is not acting. Furthermore, one of the peaks lasts for a longer time before it disappears. In order to explain these two observations, we consider $\bar{\mathcal{P}}_{cv}(\mathbf{k},t)$ and $\Sigma^{\bar{\mathcal{P}}}_{cv}(\mathbf{k},t)$, both in the lab frame and including the phase introduced in Eq.~\eqref{E:EnvelopeMovInterPol}. Clearly this phase oscillates faster for values of $\mathbf{k}$ and $t$ where the energy difference is detuned from $\hbar \omega_o$. In the Coulomb integral, Eq.~\eqref{E:FirstStepCoulomb}, the contribution from $\bar{\mathcal{P}}_{c' v'}(\mathbf{k},t)$ is smaller when the interband polarization oscillates faster; this occurs when $\bar{\mathcal{P}}_{c' v'}(\mathbf{k},t)$ has left the vicinity of the $\mathbf{k}$ points where the energy difference is close to $\hbar \omega_o$. The insets in Figs.~\ref{F:G1_CoulTerm__eps1__2bandvs14band}c and \ref{F:G1_CoulTerm__eps1__2bandvs14band}f show the shape of $|\Sigma_{cv}^{\bar{\mathcal{P}}}(\mathbf{k},t)| = |\bar{\mathfrak{S}}_{cv}(\mathbf{k}-\epsilon t \hat{\mathbf{e}}^{\text{dc}},t)|$ in the lab frame, which results in one peak lasting longer than the other in the moving frame. Consequently, the Coulomb term enhances the peaks of $|\bar{\mathfrak{P}}_{cv}(\boldsymbol \kappa,t)|$ asymmetrically, as shown in Figs.~\ref{F:G1_CoulTerm__eps1__2bandvs14band}b and \ref{F:G1_CoulTerm__eps1__2bandvs14band}e. The peak that is enhanced the most is also broadened, which is not surprising considering that the Coulomb interaction attempts to keep the electron-hole pair together, localizing its wavefunction in real space.\\


\begin{figure*}
\centering
\includegraphics[width=11.0cm]{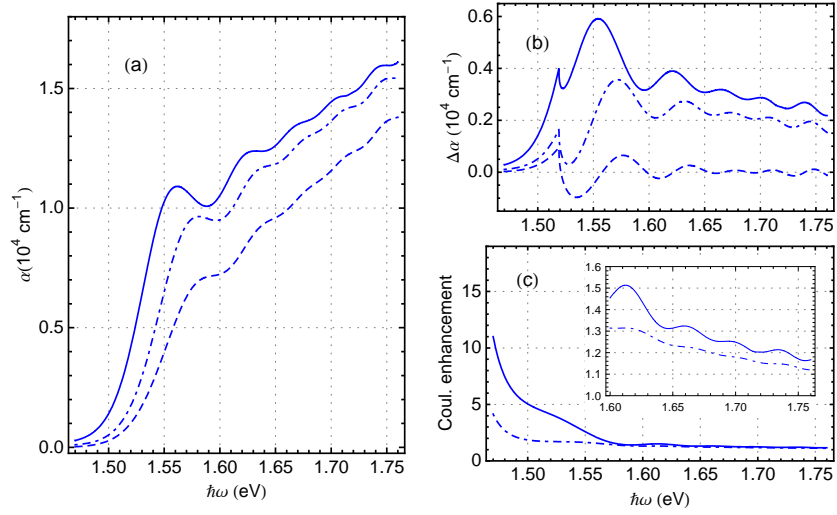}
\caption{(Color online) Effect of static screening on the Franz-Keldysh spectrum calculated in the 14-band model for a 66 kV/cm dc field oriented along the [001] direction and light polarized along the [100] direction, as in Fig.~\ref{F:eps1__x_z__Ind_Coul}. The screening length used here is 31 \AA, which is four times smaller than the radius of the exciton associated with the electron-hole pair in the effective mass approximation. For comparison we have included the cases with bare Coulomb interaction (solid blue line) and without (dashed blue line) from Fig.~\ref{F:eps1__x_z__Ind_Coul} in all the panels (a)-(c). (a) Absorption coefficient with statically screened Coulomb interaction (dash-dotted blue line). (b) Difference with respect to the zero-field absorption in the independent particle approximation for $\alpha(\omega)$ calculated with screened Coulomb interaction (dash-dotted blue line). (c) Ratio between the absorption coefficient with screened Coulomb interaction and without (dash-dotted blue line); the inset shows the energy range from 1.60 to 1.76 eV in more detail.}
\label{F:eps1__14band__Ind_Coul_Screened31Coul}
\end{figure*}

We also note that the absorption spectrum with electron-hole interaction can be modified by introducing a \emph{statically} screened potential
\begin{equation}
	U_s(\mathbf{q}) = \frac{4 \pi e^2}{\epsilon_{\text{b}}} \frac{1}{q^2 + \kappa_s^2}, \label{E:StaticallyScreenedCoulomb} 
\end{equation}
where $\kappa_s$ is the inverse of the screening length of the long-ranged bare Coulomb interaction.\cite{HaugKoch04} The screening reduces the Coulomb enhancement and shifts the peaks of the absorption coefficient to higher energies, especially near the band gap; as we would expect, both changes are in the direction of the independent particle result. We illustrate this observation in Fig.~\ref{F:eps1__14band__Ind_Coul_Screened31Coul}, where the absorption coefficient for 66 kV/cm dc field and [100] polarization in the 14-band model is plotted with and without screening. Even for arbitrarily large $\kappa_s$, we still have $U(\mathbf{q})/U_s(\mathbf{q}) \rightarrow 1$ as $q \rightarrow \infty$, so deviations of the absorption coefficient from that in the absence of the Coulomb interaction persist.\\


\begin{figure*}
\centering
\includegraphics[width=11.0cm]{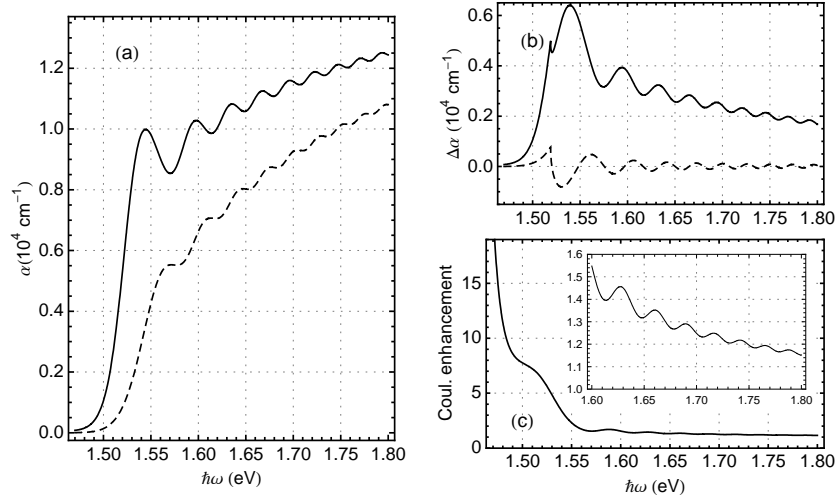}
\caption{Franz-Keldysh spectrum for a 44 kV/cm dc field calculated in the two-band model. (a) Absorption coefficient with Coulomb interaction (solid line) and without (dashed line). (b) Difference with respect to the zero-field absorption in the independent particle approximation for $\alpha(\omega)$ calculated with Coulomb interaction (solid line) and without (dashed line). (c) Ratio between the absorption coefficient with and without Coulomb interaction; the inset shows the energy range from 1.6 to 1.8 eV in more detail.}
\label{F:eps0p666__2band__Ind_Coul}
\end{figure*}

\begin{figure*}
\centering
\includegraphics[width=11.0cm]{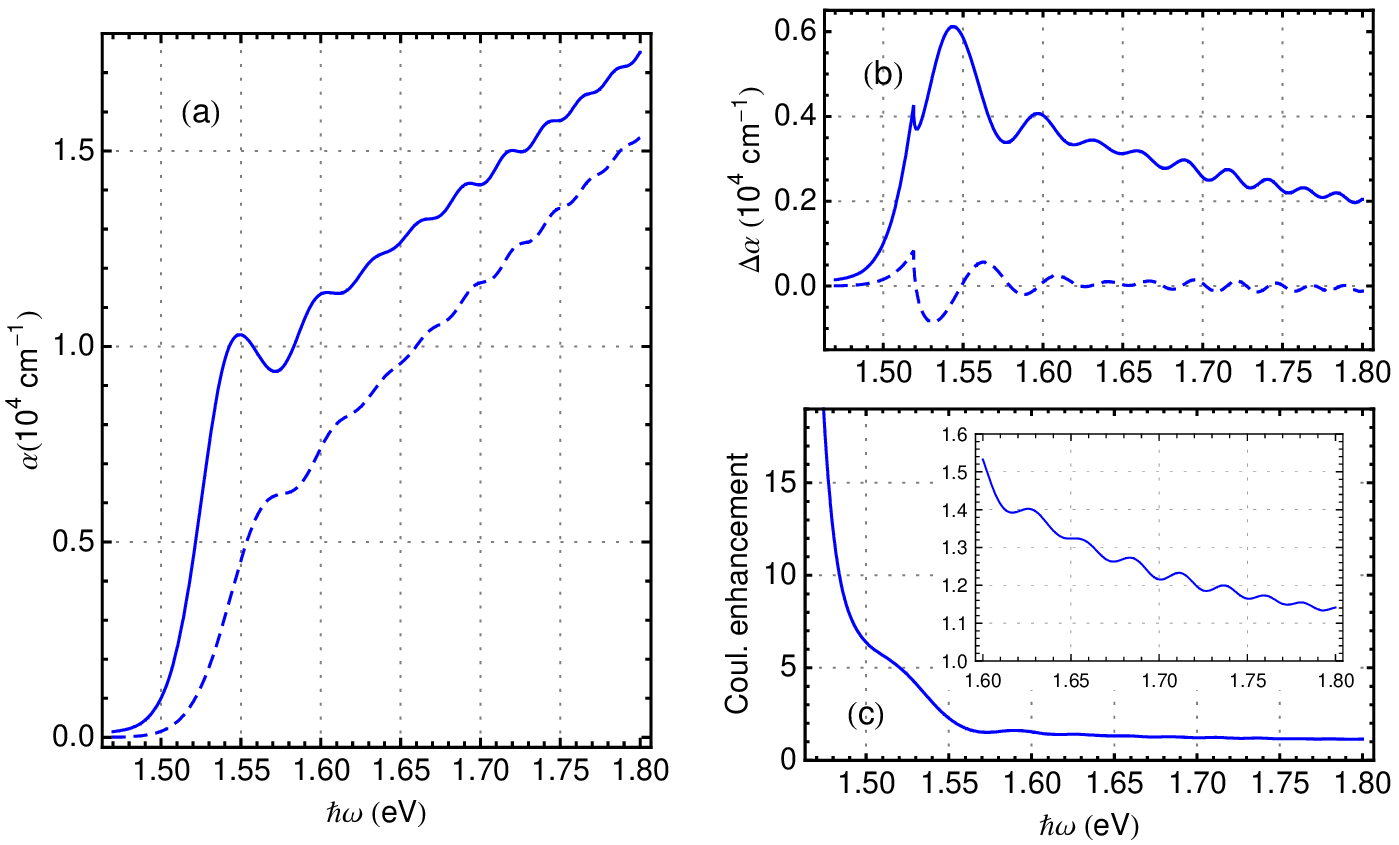}
\caption{(Color online) Franz-Keldysh spectrum for a 44 kV/cm dc field pointing  along [001] and [100] polarized light, calculated in the 14-band model. (a) Absorption coefficient with Coulomb interaction (solid blue line) and without (dashed blue line). (b) Difference with respect to the zero-field absorption in the independent particle approximation for $\alpha(\omega)$ calculated with Coulomb interaction (solid blue line) and without (dashed blue line). (c) Ratio between the absorption coefficient with and without Coulomb interaction; the inset shows the energy range from 1.6 to 1.8 eV in more detail.}
\label{F:eps0p666__14band__Ind_Coul}
\end{figure*}

Finally, we consider the effect of a weaker dc field. At the independent particle level, we expect the oscillations of the differential absorption $\Delta \alpha (\omega)$ to have a smaller amplitude and a shorter period. In the transverse configuration, the latter property allows us to observe the beating of the oscillations due to the different effective masses of the heavy and light holes, before the appearance of the `bump' associated with the split-off bands.\cite{Wahlstrand10} For these reasons, we choose the value $E^{\text{dc}}=44 \text{ kV/cm}$ for the dc field, still oriented along the [001] direction, and we set the optical polarization along the [100] direction in Figs.~\ref{F:eps0p666__2band__Ind_Coul} and \ref{F:eps0p666__14band__Ind_Coul}, where the absorption coefficient is calculated in the two-band and 14-band models, respectively. The overall behavior is very similar to the results obtained for the stronger field presented in Figs.~\ref{F:eps1__2band__Ind_Coul} and \ref{F:eps1__x_z__Ind_Coul} with the same optical polarization; however, the Coulomb enhancement increases near the band edge, roughly for energies below $1.6 \text{ eV}$ (compare Figs.~\ref{F:eps1__2band__Ind_Coul}c and \ref{F:eps0p666__2band__Ind_Coul}c for the two-band model, and Figs.~\ref{F:CoulEnhancement__eps1__x_z} and \ref{F:eps0p666__14band__Ind_Coul}c for the 14-band model). Note that the Coulomb interaction enhances the absorption at higher energies in a similar way for the two fields.\\


\section{CONCLUSION \label{S:Conclusion}} 

We have presented a theoretical approach for studying the Franz-Keldysh effect in the linear response regime. Our treatment represents an improvement on previous calculations in the independent particle approximation by including the electron-hole interaction at a Hartree-Fock level. We have kept the features associated with realistic band structure calculations from $\mathbf{k} \cdot \mathbf{p}$ models, properly accounting for the interband coupling away from the $\Gamma$ point.\cite{Enderlein75, Hader97, Wahlstrand10} We have also improved the expressions derived by Wahlstrand and Sipe by using a gauge-independent formalism that removes divergent terms proportional to $1/\omega$, such as Eq.~(51) in their work,\cite{Wahlstrand10} without relying on sum-rules;\cite{Gharamani93} this feature could be used in situations where the dc field or the optical pulse are replaced by excitations with frequencies below the optical range, for instance, in the terahertz regime.\cite{TongYi08, Schultze13} Furthermore, the calculation of the interband polarization in our approach is not only an intermediate step in finding the absorption coefficient but it has its own advantages. It illustrates the physics of the injection process and it becomes relevant in the analysis of experiments with pulsed excitation.\cite{WahlstrandCundiff10}\\


Regarding the effects of the Coulomb interaction, our calculations show that the absorption spectrum with and without Coulomb interaction are qualitatively very similar. The most noticeable effect is the enhancement of the absorption lineshape, especially near the band edge; this is accompanied by a slight displacement of the Franz-Keldysh oscillations to lower energies. The Coulomb enhancement depends on the strength of the dc field, which determines how easily the electron and hole in an exciton are pulled apart from each other; as expected, weaker dc fields lead to higher enhancement. Additionally, we studied the dependence on the orientation of the optical field fixing the dc field along the [001] direction and using parallel and transverse configurations for the optical polarization; such configurations are significantly different due to the heavy hole-light hole beat in the transverse case. This feature is preserved in the Coulomb case, but a more detailed comparison shows that the transverse configuration is favored slightly more by the Coulomb interaction (see Figs.~\ref{F:CoulEnhancement__eps1__x_z} and \ref{F:OptPolDiff__eps1__x_z}), a result that motivates a closer study of the optical polarization dependence in the Franz-Keldysh effect.\\ 

We believe that these calculations show how including both interband coupling and the Coulomb interaction lead to an advance in our understanding of the Franz-Keldysh effect. As experimental techniques are improved,\cite{Komkov08, Komkov09, WahlstrandCundiff10, Kayastha13, Ruff96} we expect that gradually more detailed features in the Franz-Keldysh effect will be observed, justifying analysis beyond the simple effective mass approximation. Finally, we point out that the framework presented here is not limited to bulk semiconductors, and can be extended to systems of different dimensionality; in addition, there are potential improvements to the treatment of the Coulomb interaction with more sophisticated self-energies.\\


\begin{acknowledgments}

We thank J. K. Wahlstrand for many valuable discussions and for providing his results and the numerical code used in his previous calculations;\cite{Wahlstrand10} we also benefitted from discussions with J. L. Cheng and M. Kira. This work was supported by the Natural Sciences and Engineering Council of Canada (NSERC). 

\end{acknowledgments}


\appendix*

\section{Derivation of the gauge-independent dynamical equation for the interband polarization \label{S:Gauge_invariance}} 

In this appendix we derive Eq.~\eqref{E:DysonEqBloch} starting with the formalism of non-equilibrium Green functions and using a basis of Wannier functions modified to obtain gauge-independent expressions.\cite{Kita06, KitaYamashita08} This derivation presents a framework to deal with more general problems than the one assumed in  Eq.~\eqref{E:DysonEqBloch}, but we show how such framework reduces to the simple case.\\

For a general non-equilibrium situation we introduce the matrix Green function
\begin{equation}
	\check{\mathcal{G}}(\mathbf{x}_1, t_1; \mathbf{x}_2, t_2) \equiv \left( \begin{array}{cc}
\mathcal{G}_{c}(\mathbf{x}_1, t_1; \mathbf{x}_2, t_2)  & \mathcal{G}^{<}(\mathbf{x}_1, t_1; \mathbf{x}_2, t_2) \\
\mathcal{G}^{>}(\mathbf{x}_1, t_1; \mathbf{x}_2, t_2)  & \mathcal{G}_{\tilde{c}}(\mathbf{x}_1, t_1; \mathbf{x}_2, t_2) \end{array} \right), \notag
\end{equation}
which contains the lesser Green function Eq.~\eqref{E:DefLesserGreen} and three more types of Green functions in real time: 
\begin{eqnarray}	
	\mathcal{G}^{>}(\mathbf{x}_1, t_1; \mathbf{x}_2, t_2) &\equiv& \frac{1}{i \hbar} \left< \psi(\mathbf{x}_1,t_1) \psi^{\dagger}(\mathbf{x}_2,t_2) \right>, \notag \\
	\mathcal{G}_{c}(\mathbf{x}_1, t_1; \mathbf{x}_2, t_2) &\equiv&  \theta(t_2-t_1) \mathcal{G}^{<}(\mathbf{x}_1, t_1; \mathbf{x}_2, t_2) \notag \\
	&+& \theta(t_1-t_2) \mathcal{G}^{>}(\mathbf{x}_1, t_1; \mathbf{x}_2, t_2), \notag \\
	\mathcal{G}_{\tilde{c}}(\mathbf{x}_1, t_1; \mathbf{x}_2, t_2) &\equiv&  \theta(t_1-t_2) \mathcal{G}^{<}(\mathbf{x}_1, t_1; \mathbf{x}_2, t_2) \notag \\
	&+& \theta(t_2-t_1) \mathcal{G}^{>}(\mathbf{x}_1, t_1; \mathbf{x}_2, t_2). \notag
\end{eqnarray}
For the last two, $\theta(t)$ denotes the Heaviside step function. The matrix Green function satisfies the Dyson equation
\begin{multline}
	\left[ i\hbar \frac{\partial}{\partial t_1} - \mathcal{H}_c(\mathbf{x}_1,t_1) \right] \check{\mathcal{G}}(\mathbf{x}_1, t_1; \mathbf{x}_2, t_2) \\
	- \int \check{\Sigma}(\mathbf{x}_1, t_1; \mathbf{x}_3, t_3) \, \check{\tau}_3 \, \check{\mathcal{G}}^{<}(\mathbf{x}_3, t_3; \mathbf{x}_2, t_2) \, d\mathbf{x}_3 dt_3 \\
	= \delta(\mathbf{x}_1-\mathbf{x}_2) \delta(t_1 - t_2)  \check{\tau}_3  \label{E:DysonLesserGreen}
\end{multline}
and its adjoint. In order to describe the interaction with electromagnetic fields, we use the minimal coupling Hamiltonian
\begin{equation}
	\mathcal{H}_c(\mathbf{x},t) \equiv \frac{1}{2m} \left( -i\hbar\nabla - \frac{e}{c} \mathbf{A}(\mathbf{x},t) \right)^2 + V(\mathbf{x}) + e A_{4}(\mathbf{x},t), \notag
\end{equation}
where $\mathbf{A}(\mathbf{x},t)$ and $A_{4}(\mathbf{x},t)$ are the vector and scalar potentials, respectively. We also include an external potential, $V(\mathbf{x})$, assumed to be the periodic potential of the crystal, and a self-energy $\check{\Sigma}(\mathbf{x}_1, t_1; \mathbf{x}_2, t_2)$, which contains the many-body effects. The third Pauli matrix,
\begin{equation}
	\check{\tau}_3 \equiv \left( \begin{array}{cc} 1 & 0 \\
	0 & -1 \end{array} \right), \notag
\end{equation}
keeps track of the correct combination of Green functions in $ \check{\mathcal{G}}(\mathbf{x}_1, t_1; \mathbf{x}_2, t_2)$.\cite{Kita06}\\

It is convenient to use a gauge-independent formalism where Eq.~\eqref{E:DysonLesserGreen} and its adjoint are formulated in terms of the electromagnetic fields instead of the potentials in $\mathcal{H}_c(\mathbf{x},t)$. Following the approach of Kita and Yamashita,\cite{KitaYamashita08} we introduce modified Wannier functions\cite{Luttinger51}
\begin{equation}
 	W'_{\boldsymbol \nu}(\mathbf{x},t) \equiv e^{i I_{\mathbf{x} \mathbf{R}}(t)} W_{\boldsymbol \nu}(\mathbf{x}), \label{E:DefModifiedWannierFunctions}
\end{equation}	
where 
\begin{equation}
	W_{\boldsymbol \nu}(\mathbf{x}) \equiv \sqrt{ \frac{\Omega_{\text{cell}}}{(2 \pi)^{3}}} \int_{\text{BZ}} \phi_{n \mathbf{k}}(\mathbf{x}) \, e^{-i \mathbf{k} \cdot \mathbf{R}} d\mathbf{k} \label{E:DefUsualWannierFunctions}
\end{equation}
are the usual Wannier functions,\cite{Wannier37} labeled by band index and lattice vector with $\boldsymbol \nu \equiv (n, \mathbf{R})$. In Eq.~\eqref{E:DefUsualWannierFunctions} we denote the volume of the unit cell by $\Omega_{\text{cell}}$ and use BZ for integration over the (first) Brillouin zone. The additional Peierls phase in Eq.~\eqref{E:DefModifiedWannierFunctions} is given by 
\begin{equation}
	I_{\mathbf{x} \mathbf{R}}(t) \equiv \frac{e}{\hbar c} \int_{\mathbf{R}}^{\mathbf{x}} \mathbf{A}(\mathbf{x}',t) \cdot d\mathbf{x}'. \notag
\end{equation}
where the integration is along the straight line from $\mathbf{R}$ to $\mathbf{x}$.\cite{Peierls33}\\

For situations where the only applied field is a uniform electric field $\mathbf{E}(t)$, we can start with a vector potential gauge such that
\begin{equation}
	A_4(\mathbf{x},t) = 0 \, \text{ and } \, \mathbf{E}(t) = -\frac{1}{c} \frac{\partial}{\partial t} \mathbf{A}(t). \notag
\end{equation} 
In this case the $W'_{\boldsymbol \nu}(\mathbf{x},t)$ are orthogonal since $\mathbf{A}(t)$ is uniform.\footnote{There are strategies to deal with the more general case where the modified Wannier functions are not orthogonal due to a non-uniform vector potential $\mathbf{A}(\mathbf{r},t)$. See Kita and Yamashita\cite{KitaYamashita08}.} In order to make the Green function in the representation of the modified Wannier functions gauge-independent, we use the two-time Peierls phase
\begin{equation}
	\mathcal{I}(\mathbf{R}_1, t_1; \mathbf{R}_2, t_2) \equiv \frac{e}{\hbar c} \frac{\mathbf{R}_1 - \mathbf{R}_2}{t_1 - t_2} \int_{t_2}^{t_1} \mathbf{A}(t) \cdot dt, \notag
\end{equation}
which is a particular case of the general definition given by Kita and Yamashita.\cite{KitaYamashita08} Thus, we introduce the gauge-independent matrix Green function
\begin{multline}
	\check{G}_{\boldsymbol \nu_1 \boldsymbol \nu_2}(t_1, t_2) \equiv e^{-i \mathcal{I}(\mathbf{R}_1, t_1; \mathbf{R}_2, t_2)} \\
	\times \int d\mathbf{x}_1 d\mathbf{x}_2 W^{\prime \ast}_{\boldsymbol \nu_1}(\mathbf{x}_1, t_1) \check{\mathcal{G}}(\mathbf{x}_1, t_1; \mathbf{x}_2, t_2) W^{\prime}_{\boldsymbol \nu_2}(\mathbf{x}_2, t_2), \label{E:ProjectionGreenModWannier}
\end{multline}
which satisfies the Dyson equation
\begin{multline}
	i \hbar \frac{\partial}{\partial t_1} \check{G}_{\boldsymbol \nu_1 \boldsymbol \nu_2}(t_1, t_2) - \sum_{\boldsymbol \nu_3} \int dt_3 \, \delta(t_1-t_3) \bigg[ \frac{\hbar \phi_{c} \delta_{\boldsymbol \nu_1 \boldsymbol \nu_3}}{t_1-t_2} \\
	+  e^{i \phi_{c}} \left( \beta_{\boldsymbol \nu_1 \boldsymbol \nu_3}(t_1)  + \check{\Sigma}_{\boldsymbol \nu_1 \boldsymbol \nu_3}(t_1) \right) \bigg] \check{G}_{\boldsymbol \nu_3 \boldsymbol \nu_2}(t_3, t_2)\\
	= \delta_{\boldsymbol \nu_1 \boldsymbol \nu_2}\delta(t_1 - t_2) \check{\tau}_3 \label{E:GaugeInvariantDyson}
\end{multline}
and its adjoint. The function $\phi_{c}$ is a gauge-independent combination of phases in a closed circuit in space-time,\cite{Kita01}
\begin{multline}
	\phi_{c} \equiv \mathcal{I}(\mathbf{R}_1, t_1; \mathbf{R}_3, t_3) \\
	+ \mathcal{I}(\mathbf{R}_3, t_3; \mathbf{R}_2, t_2) + \mathcal{I}(\mathbf{R}_2, t_2; \mathbf{R}_1, t_1). \notag
\end{multline}
Additionally in Eq.~\eqref{E:GaugeInvariantDyson}, we introduced the single-particle energy
\begin{equation}
	\beta_{\boldsymbol \nu_1 \boldsymbol \nu_2}(t) \equiv \int  W^{\ast}_{\boldsymbol \nu_1}(\mathbf{x}) \left[ \mathcal{H}_{o} - e \mathbf{E}(t) \cdot (\mathbf{x}-\mathbf{R}_2) \right] W_{\boldsymbol \nu_2}(\mathbf{x}) d\mathbf{x}, \label{E:SingleParticleEnergyWannier}
\end{equation}
where $\mathcal{H}_{o} $ is the unperturbed Hamiltonian (see Eq.~\eqref{E:DefUnperturbedHamiltonian}), and for simplicity we assumed a self-energy singular in time,
\begin{equation}
	\check{\Sigma}(\mathbf{x}_1, t_1; \mathbf{x}_2, t_2) = \check{\Sigma} (\mathbf{x}_1, \mathbf{x}_2; t_1) \delta(t_1 - t_2), \notag
\end{equation}
which was projected on the modified Wannier function basis similarly to Eq.~\eqref{E:ProjectionGreenModWannier},
\begin{multline}
	\check{\Sigma}_{\boldsymbol \nu_1 \boldsymbol \nu_2}( t) \equiv \lim_{t_1, t_2 \rightarrow t} e^{-i \mathcal{I}(\mathbf{R}_1, t_1; \mathbf{R}_2, t_2)} \\
	\times \int  W^{\prime \ast}_{\boldsymbol \nu_1}(\mathbf{x}_1,t_1) \check{\Sigma} (\mathbf{x}_1, \mathbf{x}_2; t_1) W^{\prime}_{\boldsymbol \nu_2}(\mathbf{x}_2, t_2) \, d\mathbf{x}_1 d\mathbf{x}_2. \notag
\end{multline}\\

In the Hartree-Fock approximation used here (see Eq.~\eqref{E:ExchangeSelfEnergy}) we can write 
\begin{equation}
	\check{\Sigma}_{\boldsymbol \nu_1 \boldsymbol \nu_2}( t) = \Sigma^{\text{ex}}_{\boldsymbol \nu_1 \boldsymbol \nu_2} (t) \check{\tau}_3, \notag
\end{equation}
which decouples the four elements of the matrix Green function in Eq.~\eqref{E:GaugeInvariantDyson} since it is proportional to $\check{\tau}_3$. There are further simplifications in the equal time limit because the phase factor multiplying the single-particle and self-energy terms vanishes. Thus, we take the difference between Eq.~\eqref{E:GaugeInvariantDyson} and its adjoint in the equal time limit for the lesser Green function,
\begin{multline}
	\left[ i \hbar \frac{\partial}{\partial t} + e (\mathbf{R}_1 - \mathbf{R}_2) \cdot \mathbf{E}(t) \right] G^{<}_{\boldsymbol \nu_1 \boldsymbol \nu_2}(t) \\
	- \sum_{\boldsymbol \nu_3} \bigg[ \left( \beta_{\boldsymbol \nu_1 \boldsymbol \nu_3}(t) + \Sigma^{\text{ex}}_{\boldsymbol \nu_1 \boldsymbol \nu_3} (t) \right) G^{<}_{\boldsymbol \nu_3 \boldsymbol \nu_2}(t) \\
	- G^{<}_{\boldsymbol \nu_1 \boldsymbol \nu_3}(t) \left( \beta_{\boldsymbol \nu_3 \boldsymbol \nu_2}(t) +  \Sigma^{\text{ex}}_{\boldsymbol \nu_3 \boldsymbol \nu_2} (t) \right)  \bigg] = 0, \notag
\end{multline}
where 
\begin{equation}
	G^{<}_{\boldsymbol \nu_1 \boldsymbol \nu_2}(t) \equiv \lim _{t_1, \, t_2 \rightarrow t}G^{<}_{\boldsymbol \nu_1 \boldsymbol \nu_2}(t_1, t_2).
\end{equation}\\

The single-particle energy, Eq.~\eqref{E:SingleParticleEnergyWannier}, depends on the lattice vectors only through the relative coordinate $\mathbf{R} \equiv \mathbf{R}_1 - \mathbf{R}_2$; therefore, we can write 
\begin{equation}
	\beta_{n_1 n_2}(\mathbf{R}, t) =  \int  W^{\ast}_{n_1}(\mathbf{x}-\mathbf{R}) \left[ \mathcal{H}_{o} - e \mathbf{E}(t) \cdot \mathbf{x} \right] W_{n_2}(\mathbf{x}) d\mathbf{x} \label{E:SingleParticleEnergyRelCoord}
\end{equation}
for $\beta_{\boldsymbol \nu_1 \boldsymbol \nu_2}(t)$. Note that $- e \mathbf{E}(t) \cdot \mathbf{x}$ could be replaced by $- e \mathbf{E}(t) \cdot (\mathbf{x}-\mathbf{R})$ since Wannier functions at different lattice sites are orthogonal. In Eq.~\eqref{E:SingleParticleEnergyRelCoord} we denoted the Wannier function for band $n$ and $\mathbf{R}=0$ by $W_{n}(\mathbf{x})$. Furthermore, if we assume that $G^{<}_{\boldsymbol \nu_1 \boldsymbol \nu_2}(t)$ also depends on the lattice vectors only through the relative coordinate, it can be shown that the same property holds for the self-energy $\Sigma^{\text{ex}}_{\boldsymbol \nu_1 \boldsymbol \nu_2} (t)$. Thus, we can simply write $G^{<}_{n_1  n_2}(\mathbf{R}, t)$ for $G^{<}_{\boldsymbol \nu_1 \boldsymbol \nu_2}(t)$ and
\begin{equation} 
	 \Sigma_{n_1 n_2} (\mathbf{R}, t) = i \hbar \sum_{\boldsymbol \nu_1^{\prime} \boldsymbol \nu_2^{\prime}} U_{n_1 n_2}^{\boldsymbol \nu_1^{\prime} \boldsymbol \nu_2^{\prime}}(\mathbf{R}) G^{<}_{n_1^{\prime} n_2^{\prime}}(\mathbf{R}_1^{\prime} - \mathbf{R}_2^{\prime} +\mathbf{R}, t) \label{E:SelfEnergyRelCoord}
\end{equation}
for $ \Sigma^{\text{ex}}_{\boldsymbol \nu_1 \boldsymbol \nu_2} (t)$. In Eq.~\eqref{E:SelfEnergyRelCoord} we introduced the matrix element of the Coulomb interaction in the Wannier function basis as
\begin{multline}
	U_{n_1 n_2}^{\boldsymbol \nu_1^{\prime} \boldsymbol \nu_2^{\prime}}(\mathbf{R}) \equiv \int d \mathbf{x}_1 d \mathbf{x}_2 \\
	\times W^{\ast}_{n_1}(\mathbf{x}_1) W_{\boldsymbol \nu_1^{\prime}}(\mathbf{x}_1) U(\mathbf{x}_1 - \mathbf{x}_2 + \mathbf{R}) W_{\boldsymbol \nu_2^{\prime}}^{\ast}(\mathbf{x}_2) W_{n_2}(\mathbf{x}_2). \notag
\end{multline}
At this point we move to reciprocal space using the Fourier transform of $G^{<}_{n_1  n_2}(\mathbf{R}, t)$ according to
\begin{equation}
	G^{<}_{n_1 n_2}(\mathbf{k},t) = \sum_{\mathbf{R}} G^{<}_{n_1 n_2}(\mathbf{R},t) e^{-i \mathbf{k} \cdot \mathbf{R}}, \label{E:InverseFourier}
\end{equation}
which is equivalent to the Green function introduced in Eq.~\eqref{E:GreenFuncBloch}. Similarly, the energy terms, $\beta_{n_1 n_2}(\mathbf{R}, t)$ and $\Sigma_{n_1 n_2} (\mathbf{R}, t)$, become
\begin{equation}
	\beta_{n_1 n_2}(\mathbf{k}, t) = \sum_{\mathbf{R}} \beta_{n_1 n_2}(\mathbf{R}, t) e^{-i \mathbf{k} \cdot \mathbf{R}} \notag
\end{equation}
and	
\begin{equation}	
	\Sigma_{n_1 n_2} (\mathbf{k}, t) = \sum_{\mathbf{R}} \Sigma_{n_1 n_2} (\mathbf{R}, t) e^{-i \mathbf{k} \cdot \mathbf{R}}. \notag
\end{equation}
Using Eqs.~\eqref{E:DefUsualWannierFunctions} and \eqref{E:SelfEnergyRelCoord} we can write the matrix for the self-energy, whose elements are given by $\Sigma_{n_1 n_2} (\mathbf{k}, t)$, as
\begin{multline}
	\Sigma(\mathbf{k},t) = i\hbar \int \frac{d\mathbf{q}}{2 \pi^3} U(\mathbf{q}) \\
	\times \sum_{\mathbf{R}_1^{\prime} \mathbf{R}_2^{\prime}} \mathcal{O}(\mathbf{R}_1^{\prime}, \mathbf{q}) G^{<}(\mathbf{k}-\mathbf{q},t) e^{i (\mathbf{k}-\mathbf{q}) (\mathbf{R}_1^{\prime}-\mathbf{R}_2^{\prime})} \mathcal{O}^{\dagger}(\mathbf{R}_2^{\prime}, \mathbf{q}), \notag
\end{multline}
where $\mathcal{O}(\mathbf{R}, \mathbf{q})$ is a matrix in the band indices with elements
\begin{equation}
	\mathcal{O}_{n_1 n_2}(\mathbf{R}, \mathbf{q}) = \frac{\Omega_{\text{cell}}}{(2 \pi)^3} \int_{\text{BZ}} \Delta_{n_1 n_2}(\mathbf{k}, \mathbf{k} - \mathbf{q}) e^{-i(\mathbf{k} - \mathbf{q}) \cdot \mathbf{R}} d \mathbf{k}, \notag
\end{equation}
which contains the overlap matrix defined in Eq.~\eqref{E:OverlapBlochCoulomb}; this expression for the self-energy reduces to Eq.~\eqref{E:ExchangeSelfEnergyBloch}. After these transformations we finally arrive at Eq.~\eqref{E:DysonEqBloch}.


%

\end{document}